\documentclass[usenatbib]{mn2e}

\usepackage{graphicx}
\usepackage{aas_macros}
\usepackage{amsfonts,amssymb}
\usepackage{color}

\newcommand{\lya}{Ly$\alpha$}

\newcommand{\hi}{H~{\sc i}}
\newcommand{\oi}{O~{\sc i}}
\newcommand{\cii}{C~{\sc ii}}

\newcommand{\civ}{C~{\sc iv}}
\newcommand{\siii}{Si~{\sc ii}}

\newcommand{\siiv}{Si~{\sc iv}}
\newcommand{\nv}{N~{\sc v}}

\newcommand{\mgii}{Mg~{\sc ii}}
\newcommand{\feii}{Fe~{\sc ii}}

\newcommand{\hei}{He~{\sc i}}
\newcommand{\heii}{He~{\sc ii}}
\newcommand{\heiii}{He~{\sc iii}}

\newcommand{\kms}{km~s$^{-1}$}

\newcommand{\taueff}{$\tau_{\rm eff}$}

\newcommand{\mac}{$\left< |\kappa| \right>$}
\newcommand{\logmac}{$log{\left< |\kappa| \right>}$}
\newcommand{\To}{$T_{0}$}
\newcommand{\Tdelta}{$T(\bar{\Delta})$}
\newcommand{\g}{$\gamma$}
\newcommand{\Fr}{$F^{R10}$}
\newcommand{\hinv}{$h^{-1}$}

\title[IGM Temperatures over $2 < z < 5$]{Detection of Extended \heii\ Reionization in the Temperature Evolution of the Intergalactic Medium\thanks{A part of the observations were made at the W.M. Keck Observatory which is operated as a scientific partnership between the California Institute of Technology and the University of California; it was made possible by the generous support of the W.M. Keck Foundation.  This paper also includes data gathered with the 6.5 meter Magellan Telescopes located at Las Campanas Observatory, Chile.}}

\author[G. D. Becker et al.] 
 {George D. Becker$^{1}$\thanks{E-mail: gdb@ast.cam.ac.uk}, 
  James S. Bolton$^{2}$, 
  Martin G. Haehnelt$^{1}$,\newauthor 
  Wallace L. W. Sargent$^{3}$ \\ 
  $^1$ Kavli Institute for Cosmology and Institute of Astronomy, Madingley Rd, Cambridge, CB3 0HA, UK \\
  $^2$ School of Physics, University of Melbourne, Parkville, VIC 3010, Australia \\
  $^3$ California Institute of Technology, 1201 E California Blvd, Pasadena, CA, 91125, USA \\}

\begin{document}

\date{Accepted 2010 August 11.}

\maketitle

\begin{abstract}

We present new measurements of the temperature of the intergalactic
medium (IGM) derived from the \lya\ forest over $2.0 \le z \le 4.8$.
The small-scale structure in the forest of 61 high-resolution QSO
spectra is quantified using a new statistic, the curvature, and the
conversion to temperature calibrated using a suite of hydrodynamic
simulations.  At each redshift we focus on obtaining the
temperature at an optimal overdensity probed by the \lya\ forest,
\Tdelta, where the temperature is nearly a one-to-one function of the
curvature regardless of the slope of the temperature-density relation.
The median 2$\sigma$ statistical uncertainty in these measurements is
8 per cent, though there may be comparable systematic errors due to
the unknown amount of Jeans smoothing in the IGM.  We use our \Tdelta\
results to infer the temperature at the mean density, \To.  Even for a
maximally steep temperature-density relation, \To\ must increase from
$\sim$8\,000\,K at $z \simeq 4.4$ to $\gtrsim$12\,000\,K at $z \simeq
2.8$.  This increase is not consistent with the monotonic 
decline in \To\ expected in the absence of \heii\ reionization.  We 
therefore interpret the observed rise in temperature as evidence of
\heii\ reionization beginning at $z \gtrsim 4.4$.  The evolution of
\To\ is consistent with an end to \heii\ reionization at $z \sim 3$,
as suggested by opacity measurements of the \heii\ \lya\ forest,
although the redshift at which \To\ peaks will depend somewhat on the
evolution of the temperature-density relation.  These new temperature
measurements suggest that the  heat input 
due to the reionization of \heii\ dominates the thermal balance of the
IGM over an extended period with $\Delta z \gtrsim 1$.   

\end{abstract}

\begin{keywords}
  intergalactic medium - quasars: absorption lines - cosmology:
  observations - cosmology: early Universe
\end{keywords}

\section{Introduction}

Cosmic reionization represents one of the most dramatic feedback processes occurring between luminous sources and the intergalactic medium (IGM).  At some point before redshift six, nearly every hydrogen atom in the Universe was ionized, as evidenced by the transmission of \hi\ \lya\ photons in QSO spectra up to $z \sim 6$ \citep{becker2001, fan2002, fan2006b}.  Similarly, helium in the IGM is known to be fully ionized by $z \sim 2.7$ based on observed transmission in the \heii\ \lya\ forest at $z < 3$ \citep[for a summary of observations, see][]{dixon2009}.   Understanding how and when these reionization events occurred is fundamental to understanding the nature of the earliest galaxies and AGN, as well in assessing the impact of reionization on the evolution of luminous sources up to the present day. 

One of the main impacts of hydrogen and helium reionization is on the thermal state of the IGM.  This is particularly true at low densities, where photoionization heating is believed to be the dominant heating process.   When the bulk of the hydrogen and helium in the IGM becomes ionized, the heat input is expected to significantly increase the gas temperature.  The thermal history of the IGM will reflect not only the timing and duration of reionization, but also the nature of the ionizing sources and the manner in which ionized bubbles grow and overlap \citep{huignedin1997, abel1999, gnedin2000a, huihaiman2003, gleser2005, paschos2007, tittley2007, furoh2008b, boltonohfur2009a, mcquinn2009a}.   At the mean density, the characteristic signature of reionization is expected to be a marked heating followed by a relatively rapid cooling, the latter due to adiabatic expansion and (at $z > 6$) Compton cooling \citep{me1994,huignedin1997}.  If helium reionizes later than hydrogen, as expected if helium is fully reionized by bright QSOs, then the temperature evolution will contain two peaks.  Following reionization, the thermal evolution will asymptotically approach a gradual cooling state, in which adiabatic expansion is nearly balanced by photoionization heating from \hi, \hei, and/or \heii\ \citep{huignedin1997,huihaiman2003}.  In short, the thermal history of the IGM reflects both the output of luminous sources and the growth of large-scale structure.

At $z < 6$, the most readily observable impact of the IGM temperature is on the small-scale structure of the \hi\ \lya\ forest.  The widths of absorption features are set by thermal broadening, Hubble broadening due to the change in redshift across a filament or void, peculiar velocities, and the column densities of the absorbers.  By using numerical simulations to capture the large-scale structure and bulk motions of the IGM, therefore, the temperature can be extracted by analyzing the amount of small-scale structure in the forest.  Previous efforts to to do this have taken two main approaches.  The first has been to measure the distribution of Doppler parameters of individual absorption features \citep{schaye2000, ricotti2000, mcdonald2001b}.  The widths of the narrowest features are expected to be dominated by thermal broadening, and so can be used to infer a temperature.  Alternatively, the small-scale structure has been quantified across the entire forest using either the flux power spectrum \citep{zaldarriaga2001} or wavelet decomposition \citep{theuns2000b, zaldarriaga2002, theuns2002b, theuns2002a}.  

Obtaining reliable measurements of the IGM temperature, however, has remained a considerable challenge.    \citet{schaye2000} found tentative evidence for an increase in the IGM temperature consistent with \heii\ reionization near $z \sim 3$.  In constrast, \citet{mcdonald2001b} found the IGM temperature to be roughly constant over $z \sim 2-4$ using a very similar data set but a somewhat different line fitting method.  \citet{lidz2009} recently conducted the most careful analysis to date using wavelets.  Their study included a large data set and paid close attention to systematic effects, yet the uncertainties in the temperature measurements remained substantial.

This work aims to produce robust IGM temperature measurements over $2 < z < 5$ using a method that is new in two ways.  First, we introduce a new statistic, the curvature, that is well suited to quantifying the small-scale structure of the \lya\ forest.  The curvature is essentially the second derivative of the flux with respect to wavelength or relative velocity, and has a number of advantages for this type of work.  It is straightforward to compute and can be evaluated on a pixel-by-pixel basis.  It also does not require the forest to be decomposed into individual lines, and so can be applied at $z > 4$ where absorption features start to become strongly blended.  Second, we focus our efforts on accurately measuring the temperature at a well-chosen overdensity, rather than attempting to constrain the entire temperature-density relation.  The temperature-density relation is typically described by a power law, $T(\Delta) = T_{0}\Delta^{\gamma-1}$, where \To\ is the temperature at the mean density and $\Delta = \rho / \langle \rho \rangle$.  We show that, at each redshift, an optimal overdensity, $\bar{\Delta}$, can be identified where \Tdelta\ is a one-to-one function of the mean absolute curvature, regardless of \g.  This allows us to normalize the temperature-density relation with high precision, providing a first step towards a full characterization of the thermal state of the IGM.

The remainder of the paper is organized as follows:  We introduce the data in Section~\ref{sec:data}, and the simulations to which the data are compared in Section~\ref{sec:sims}.  We introduce the curvature statistic and describe our method for translating the curvature into a temperature measurement  in Section~\ref{sec:method}.  Here we also discuss our strategies for dealing with sources of potential systematic error, as well as examine the potential impact of the thermal history of the IGM on measuring instantaneous temperatures.  Our results are presented in Section~\ref{sec:results}.  These include the temperature at the optimal overdensity, as well as the temperature at the mean density for different values of \g.  The implications for \heii\ reionization are discussed in Section~\ref{sec:implications}.  Finally, we summarize our conclusions in Section~\ref{sec:summary}.  The numerical convergence of our simulations is examined in the Appendix.  Comoving distances are used throughout the paper.

\section{The Data}\label{sec:data}

\begin{table}
   \caption{List of QSOs analyzed in this work.  Columns give the QSO name, the emission redshift, the instrument with which the spectrum was taken, the total exposure time across all settings, and the mean noise in the normalized flux per 2.1\,\kms\  among those sections of the spectra that were included in the analysis.  In order to be included, a 10\,\hinv\,Mpc section was required to have a mean noise level less than 0.06. Objects for which `HIRES' is marked with an $(^{*})$ were observed at least partially with the upgraded detector.\vspace{-0.1in}}
   \label{tab:qsos}
   \begin{center}
   \begin{tabular*}{8.4cm}{@{\extracolsep{\fill}}lcccc}
   \hline
    QSO  &  $z_{\rm em}$  &  Instrument  &  $t_{\rm exp}$  &  Noise \\
   \hline
    PKS~0421$+$019   &  2.05  &      HIRES        &  3.3   &  0.057  \\
      Q~1225$+$3145  &  2.20  &  \ \,HIRES$^{*}$  &  8.7   &  0.015  \\
     HS~1103$+$6416  &  2.21  &  \ \,HIRES$^{*}$  &  1.5   &  0.028  \\
      Q~1521$+$5202  &  2.22  &  \ \,HIRES$^{*}$  &  2.0   &  0.023  \\
    PKS~0237$-$23    &  2.24  &      HIRES        &  6.7   &  0.047  \\
     HS~1626$+$6433  &  2.32  &  \ \,HIRES$^{*}$  &  13.5  &  0.027  \\
     HE~1122$-$1648  &  2.42  &  \ \,HIRES$^{*}$  &  5.0   &  0.020  \\
      Q~1623$+$2653  &  2.53  &      HIRES        &  3.3   &  0.048  \\
      Q~1017$-$2046  &  2.55  &  \ \,HIRES$^{*}$  &  4.0   &  0.027  \\
     HS~1603$+$3820  &  2.56  &  \ \,HIRES$^{*}$  &  14.8  &  0.010  \\
      Q~2206$-$1958  &  2.57  &      HIRES        &  8.3   &  0.033  \\
      Q~2343$+$1232  &  2.58  &      HIRES        &  11.7  &  0.046  \\
     HE~1347$-$2457  &  2.59  &  \ \,HIRES$^{*}$  &  4.5   &  0.020  \\
     HS~0818$+$3117  &  2.62  &  \ \,HIRES$^{*}$  &  3.0   &  0.033  \\
      Q~1310$+$4254A &  2.63  &  \ \,HIRES$^{*}$  &  2.7   &  0.043  \\
      Q~1009$+$2956  &  2.65  &  \ \,HIRES$^{*}$  &  11.5  &  0.009  \\
      Q~1442$+$2931  &  2.66  &  \ \,HIRES$^{*}$  &  18.2  &  0.015  \\
      Q~0100$+$1300  &  2.71  &      HIRES        &  5.8   &  0.026  \\
     HS~1700$+$6416  &  2.74  &  \ \,HIRES$^{*}$  &  17.8  &  0.009  \\
      Q~2344$+$1228  &  2.79  &      HIRES        &  4.2   &  0.050  \\
     HS~1549$+$1919  &  2.84  &  \ \,HIRES$^{*}$  &  17.8  &  0.005  \\
     HS~0119$+$1432  &  2.87  &      HIRES        &  9.2   &  0.031  \\
     HS~1132$+$2243  &  2.88  &      HIRES        &  3.3   &  0.056  \\
      Q~1511$+$0907  &  2.89  &      HIRES        &  4.2   &  0.051  \\
      Q~1107$+$4847  &  2.98  &      HIRES        &  8.3   &  0.029  \\
     HS~1437$+$3007  &  2.98  &      HIRES        &  8.7   &  0.046  \\
      Q~2231$-$0015  &  3.02  &      HIRES        &  4.2   &  0.044  \\
     HS~1946$+$7658  &  3.07  &      HIRES        &  7.5   &  0.020  \\
     HE~0940$-$1050  &  3.08  &      HIRES        &  3.3   &  0.039  \\
      Q~0449$-$1325  &  3.10  &      HIRES        &  14.2  &  0.044  \\
     HS~1011$+$4315  &  3.14  &      HIRES        &  6.7   &  0.035  \\
      Q~1140$+$3508  &  3.16  &      HIRES        &  5.8   &  0.026  \\
      Q~0636$+$6801  &  3.17  &      HIRES        &  10.8  &  0.018  \\
     HS~1425$+$6039  &  3.18  &      HIRES        &  10.8  &  0.012  \\
     HS~0741$+$4741  &  3.23  &      HIRES        &  8.3   &  0.024  \\
     HS~0757$+$5218  &  3.25  &      HIRES        &  5.8   &  0.024  \\
      Q~0428$-$1342  &  3.25  &      HIRES        &  4.2   &  0.056  \\
      Q~0302$-$0019  &  3.28  &      HIRES        &  7.5   &  0.035  \\
      Q~0956$+$1217  &  3.31  &      HIRES        &  3.3   &  0.042  \\
      Q~0642$+$4454  &  3.40  &      HIRES        &  5.8   &  0.048  \\
      Q~0930$+$2858  &  3.44  &      HIRES        &  6.7   &  0.052  \\
      Q~1422$+$2309  &  3.63  &      HIRES        &  9.2   &  0.019  \\
      Q~0055$-$2659  &  3.65  &      HIRES        &  6.7   &  0.044  \\     
   APM~08279$+$5255  &  3.91  &      HIRES        &  7.5   &  0.012  \\
    BRI~0241$-$0146  &  4.07  &      HIRES        &  10.0  &  0.044  \\
    PSS~1646$+$5514  &  4.10  &      HIRES        &  15.8  &  0.017  \\
      Q~0000$-$2620  &  4.13  &      HIRES        &  3.3   &  0.039  \\
    PSS~1057$+$4555  &  4.15  &      HIRES        &  8.3   &  0.024  \\
    PSS~0209$+$0517  &  4.18  &      HIRES        &  14.2  &  0.022  \\
    PSS~0248$+$1802  &  4.44  &      HIRES        &  12.5  &  0.046  \\
     BR~0418$-$5723  &  4.48  &      MIKE         &  15.8  &  0.054  \\
     BR~0714$-$6449  &  4.49  &      MIKE         &  6.7   &  0.052  \\
     BR~2237$-$0607  &  4.56  &      HIRES        &  10.8  &  0.030  \\
     BR~0353$-$3820  &  4.59  &      MIKE         &  9.2   &  0.032  \\
   SDSS~2147$-$0838  &  4.59  &      MIKE         &  8.3   &  0.050  \\
     BR~1202$-$0725  &  4.69  &  \ \,HIRES$^{*}$  &  9.2   &  0.038  \\
   SDSS~0011$+$1446  &  4.95  &  \ \,HIRES$^{*}$  &  6.7   &  0.026  \\
   \hline
   \end{tabular*}
   \end{center}
\end{table}

\addtocounter{table}{-1}

\begin{table}
   \caption{-- Continued \vspace{-0.1in}}
   \begin{center}
   \begin{tabular*}{8.4cm}{@{\extracolsep{\fill}}lcccc}
   \hline
    QSO  &  $z_{\rm em}$  &  Instrument  &  $t_{\rm exp}$  &  Noise \\
   \hline
   SDSS~0040$-$0915  &  4.98  &      MIKE         &  8.3   &  0.054  \\
   SDSS~0915$+$4924  &  5.20  &  \ \,HIRES$^{*}$  &  10.8  &  0.058  \\
   SDSS~1659$+$2709  &  5.32  &  \ \,HIRES$^{*}$  &  11.7  &  0.041  \\
   SDSS~0836$+$0054  &  5.80  &  \ \,HIRES$^{*}$  &  21.7  &  0.058  \\
   \hline
   \end{tabular*}
   \end{center}
\end{table}

Our sample includes 61 QSOs spanning emission redshifts $2.1 \le z_{\rm em} \le 5.8$.  The objects are listed in Table~\ref{tab:qsos}.  High-resolution spectra were obtained primarily with the HIRES spectrograph \citep{vogt1994} on Keck I.  Observations were made with the original CCD detector between 1993 and 2004, and with the upgraded detector between 2005 and 2008.  Additional spectra of QSOs at $z \ge 4.5$ were taken with the MIKE spectrograph on Magellan \citep{bernstein2002} between 2006 and 2008.  For QSOs at $z < 4.7$ taken with HIRES, the raw data were processed and the extracted spectra combined using the MAKEE reduction package.  At $z > 4.7$, the HIRES data were reduced using a custom set of IDL routines, as described in \citet{becker2006}.  The IDL routines were designed to handle faint targets, and include optimal sky subtraction \citep{kelson2003} as well as simultaneous optimal extraction of the 1D spectrum from all exposures.  The MIKE data were reduced using a similar custom software package.  

The resolution of the HIRES data is 6.7~\kms\ (FWHM).  The majority of
the HIRES data were binned using pixels with a constant velocity width
of 2.1~\kms, although a subset of the spectra at $z \le 2.8$ were
binned using constant 0.03~\AA\ pixels.  The resolution of the MIKE
spectra is 13.6~\kms.  For these spectra, 5.0~\kms\  binned pixels
were used.  For both instruments the resolution is sufficient to
resolve typical \lya\ forest lines, which have Doppler parameters $b
\gtrsim 10$~\kms, or ${\rm FWHM} \gtrsim 24$~\kms.  When making
comparisons with the simulations, we smooth and resample the
simulations to match the resolution and binning of each QSO spectrum
individually.  Provided that adequate sampling is used, we found the
curvature statistic (Section~\ref{sec:method}) does not depend on the
choice of binning, although it does depend on the resolution.  As
described below, the curvature is measured in sections spanning
10\,$h^{-1}$\,Mpc sections.  Only sections with a mean $rms$
uncertainty in the normalized flux of 0.06 per 2.1~\kms\
bin\footnote{This is a typical pixel size for the HIRES data.  For
  MIKE data, the noise threshold per 5.0\,\kms\ pixel was $0.06 \times
  \sqrt{2.1/5.0}$.} were included in the final analysis.  Initial
continua were fit using slowly-varying spline profiles.  Real data are 
difficult to continuum fit accurately, however, 
and we discuss our strategy for avoiding errors due to continuum
fitting 
in Section~\ref{sec:continuum}.

\section{Hydrodynamical simulations}\label{sec:sims}

In order to obtain temperature constraints from the \lya\ forest, we
have used
detailed simulated spectra for calibrating and interpreting
our observational measurements.  We performed a
large set of high resolution, fully hydrodynamical simulations which
span a wide variety of possible thermal histories.  The simulations
were run using the parallel Tree-SPH code {\small GADGET-3},
which is an updated version of the publicly available code {\small
  GADGET-2} \citep{springel2005}, and are summarized in
Table~\ref{tab:sims}.

The fiducial simulation volume is a $10\,h^{-1}$\,Mpc periodic box
containing $2 \times 512^{3}$ gas and dark matter particles.  This
resolution is chosen specifically to resolve the \lya\ forest at high
redshift \citep{boltonbecker2009}.  The simulations were all started
at $z=99$, with initial conditions generated using the transfer
function of \citet{eisensteinhu1999}.  The cosmological parameters are
$\Omega_{\rm m}=0.26$, $\Omega_{\Lambda}=0.74$, $\Omega_{\rm
  b}h^{2}=0.023$, $h=0.72$, $\sigma_{8}=0.80$, $n_{\rm s}=0.96$,
consistent with recent studies of the cosmic microwave background
\citep{reichardt2009,jarosik2010}.  The IGM is assumed to be of
primordial composition with a helium fraction by mass of $Y=0.24$
\citep{olive2004}.  The gravitational softening length was
set to $1/30^{\rm th}$ of the mean linear interparticle spacing and
star formation was included using a simplified prescription which
converts all gas particles with overdensity $\Delta = \rho/\langle
\rho \rangle > 10^{3}$ and temperature $T<10^{5}\rm~K$ into
collisionless stars.  A total of 17 outputs at different redshifts are
obtained from each simulation, spanning the range $1.831\le z \le
6.010$.

\begin{figure}
   \begin{center}
   \includegraphics[width=0.40\textwidth]{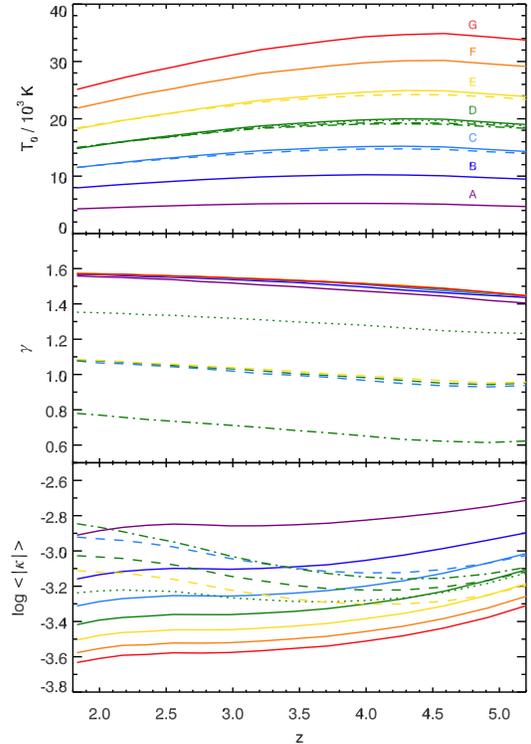}
   \vspace{-0.05in}
   \caption{Thermal properties of the the main simulations used in this work.  Lines are color coded according to \To.  Line styles denote different \g's -- {\it solid}: $\gamma \sim 1.5$, {\it dotted}: $\gamma \sim 1.3$, {\it dashed}: $\gamma \sim 1.0$, {\it dot-dashed}:  $\gamma \sim 0.7$.  Top panel: Temperature at the mean density as a function of redshift.  Letters correspond to the first letter of the simulation name in Table~\ref{tab:sims}.  Middle panel:  Slope$ + 1$ of the temperature-density relation, where $T(\Delta) = T_{0} \Delta^{\gamma-1}$.  Bottom panel: The curvature statistic computed as described in Section~\ref{sec:method}.  Here, the curvature has been calculated directly from the noise-free spectra.}
   \label{fig:sim_grid}
   \end{center}
\end{figure}

The gas in the simulations is assumed to be optically thin and in
ionization equilibrium with a spatially uniform ultraviolet background
(UVB).  The UVB corresponds to the galaxies and quasars emission model
of \citet{hm2001}.  Hydrogen is reionized at $z=9$ and gas with
$\Delta \la 10$ subsequently follows a tight power-law
temperature-density relation, $T=T_{0}\Delta^{\gamma-1}$, where
$T_{0}$ is the temperature of the IGM at mean density
\citep{huignedin1997,valageas2002}.  In order to explore a variety of
thermal histories, we rescale the \citet{hm2001} photo-heating
rates by different constants in the models.  In each simulation we
assume $\epsilon_{i}=\zeta \Delta^{\xi}\epsilon_{i}^{\rm HM01}$, where
$\epsilon_{\rm i}^{\rm HM01}$ are the \citet{hm2001}
photo-heating rates for species $i=[{\rm H \,\scriptstyle I},{\rm He
    \,\scriptstyle I}, {\rm He \,\scriptstyle II}]$ and $\zeta$, $\xi$
are constants listed in Table~\ref{tab:sims}.  Our fiducial models (A15-G15) use $\xi = 0.0$, which produces a temperature-density relation with $\gamma \sim 1.5$ at $z=3$ (Table~\ref{tab:sims}). A selection of other
models include a density-dependent rescaling of the photo-heating
rates to explore the effect of the relationship between temperature
and gas density on our results.  In two further models, T15slow and T15fast, we use a
density-independent scaling which varies with redshift.  These
simulations have integrated thermal histories which are significantly
different from that of the HM01 reference model.  We use these models to
test the impact of Jeans smoothing on our results in Section~\ref{sec:thermal_history}.  Finally, we also perform four simulations, R1, R2, R3 and R4, which use
same thermal history as model C15 but have different box sizes and
mass resolutions.  These simulations are used to test the numerical
convergence of our results, which is discussed in the appendix at the
end of this paper.  The values of $T_{0}$ and $\gamma$ in the
hydrodynamical simulations at $z=3$ are listed in columns 7 and 8 of
Table~\ref{tab:sims}, and the thermal histories as a function of
redshift for all simulations except the T and R series
are shown in Figure~\ref{fig:sim_grid}.

We construct synthetic \lya\ spectra at each redshift using the output
of our hydrodynamical simulations \citep[e.g.,][]{theuns1998}.  The
spectra are then modified in post-processing to match the
characteristics of the observational data as closely as possible.  We
now turn to discuss this aspect of our simulations in more detail, and
introduce our method for obtaining our IGM temperature constraints.

\begin{table*}
   \caption{Simulation runs used in this work.  Columns give the simulation name, the box size, the total number of particles, the gas particle mass, the scaling factors for the UVB photo-heating rates (see text for details), the median volume-weighted gas temperature at the mean density at $z=3$, the slope$ + 1$ of the temperature-density relation at $z=3$, where $T(\Delta) = T_{0}\Delta^{\gamma-1}$, and notes on the simulations.}
   \label{tab:sims}
   \begin{minipage}{\textwidth}
   \begin{center}
   \begin{tabular*}{\textwidth}{@{\extracolsep{\fill}}lcccccccl}
   \hline
   Model   & $L~[h^{-1}\rm\, Mpc]$ & Particles & $M_{\rm gas}~[h^{-1}M_{\odot}]$ & 
   $\zeta$ & $\xi$ & $T_{0}^{z=3}$ [K] & $\gamma^{z=3}$ & Notes \\
   \hline
    A15     &  10  &  $2\times 512^{3}$  &  $9.2\times 10^{4}$  &  0.30   &   0.00  &  $ 5\,100$  &  1.52  &  Fiducial $T$-$\Delta$ relation  \\
    B15     &  10  &  $2\times 512^{3}$  &  $9.2\times 10^{4}$  &  0.80   &   0.00  &  $ 9\,600$  &  1.54  &  Fiducial $T$-$\Delta$ relation  \\
    C15     &  10  &  $2\times 512^{3}$  &  $9.2\times 10^{4}$  &  1.45   &   0.00  &  $14\,000$  &  1.54  &  Fiducial $T$-$\Delta$ relation  \\
    C10     &  10  &  $2\times 512^{3}$  &  $9.2\times 10^{4}$  &  1.45   &   -1.00  &  $13\,700$  &  1.02  &  ``Isothermal'' $T$-$\Delta$ relation  \\
    D15     &  10  &  $2\times 512^{3}$  &  $9.2\times 10^{4}$  &  2.20   &   0.00  &  $18\,200$  &  1.55  &  Fiducial $T$-$\Delta$ relation  \\
    D13     &  10  &  $2\times 512^{3}$  &  $9.2\times 10^{4}$  &  2.20   &   -0.45  &  $18\,100$  &  1.32  &  Flattened $T$-$\Delta$ relation  \\
    D10     &  10  &  $2\times 512^{3}$  &  $9.2\times 10^{4}$  &  2.20   &   -1.00  &  $18\,000$  &  1.03  &  ``Isothermal'' $T$-$\Delta$ relation  \\
    D07     &  10  &  $2\times 512^{3}$  &  $9.2\times 10^{4}$  &  2.20   &   -1.60  &  $17\,900$  &  0.71  &  Inverted $T$-$\Delta$ relation  \\
    E15     &  10  &  $2\times 512^{3}$  &  $9.2\times 10^{4}$  &  3.10   &   0.00  &  $22\,500$  &  1.55  &  Fiducial $T$-$\Delta$ relation  \\
    E10     &  10  &  $2\times 512^{3}$  &  $9.2\times 10^{4}$  &  3.10   &   -1.00  &  $22\,200$  &  1.04  &  ``Isothermal'' $T$-$\Delta$ relation  \\
    F15     &  10  &  $2\times 512^{3}$  &  $9.2\times 10^{4}$  &  4.20   &   0.00  &  $27\,000$  &  1.55  &  Fiducial $T$-$\Delta$ relation  \\
    G15     &  10  &  $2\times 512^{3}$  &  $9.2\times 10^{4}$  &  5.30   &   0.00  &  $31\,000$  &  1.55  &  Fiducial $T$-$\Delta$ relation  \\
    T15slow &  10  &  $2\times 512^{3}$  &  $9.2\times 10^{4}$  &  Varied &   0.00  &  $18\,200$  &  1.53  &  Test run with gradual temperature increase \\
    T15fast &  10  &  $2\times 512^{3}$  &  $9.2\times 10^{4}$  &  Varied &   0.00  &  $18\,600$  &  1.54  &  Test run with rapid temperature increase \\
    R1      &  10  &  $2\times 256^{3}$  &  $7.4\times 10^{5}$  &  1.45   &   0.00  &  $13\,700$  &  1.02  &  Mass resolution convergence  \\
    R2      &  10  &  $2\times 128^{3}$  &  $5.9\times 10^{6}$  &  1.45   &   0.00  &  $13\,700$  &  1.02  &  Mass resolution/box size convergence  \\
    R3      &  20  &  $2\times 256^{3}$  &  $5.9\times 10^{6}$  &  1.45   &   0.00  &  $14\,000$  &  1.54  &  Box size convergence  \\
    R4      &  40  &  $2\times 512^{3}$  &  $5.9\times 10^{6}$  &  1.45   &   0.00  &  $14\,000$  &  1.54  &  Box size convergence  \\
  \hline
   \end{tabular*}
   \end{center}
   \end{minipage}
\end{table*}

\section{Method}\label{sec:method}

\subsection{The curvature statistic}

We introduce a new statistic, the curvature, to efficiently measure the amount of small-scale structure in the \lya\ forest.  The curvature, $\kappa$, is conventionally defined as

\begin{equation}
   \kappa \equiv \frac{F''}{\left[ 1 + (F')^{2} \right]^{3/2}} \, .
   \label{eq:curvature}
\end{equation}

\noindent We take the derivative of the flux with respect to the velocity separation between pixels, which is measured in \kms.  In practice, the change in flux per \kms\ is generally small, and so the denominator is always close to unity.  The curvature is therefore essentially equal to the second derivative of the flux.

The curvature has several advantages over other measurements of the small-scale structure.  It is easy to compute, and can be evaluated on a pixel-by-pixel basis.  It is also a continuous function, and does not require the forest to be deconstructed into individual absorption features.  This allows the curvature to be easily calculated at high redshifts ($z \gtrsim 4$), where high levels of absorption make it difficult to identify individual absorption lines.
 
\begin{figure*}
   \centering
   \begin{minipage}{\textwidth}
   \begin{center}
   \includegraphics[width=0.75\textwidth]{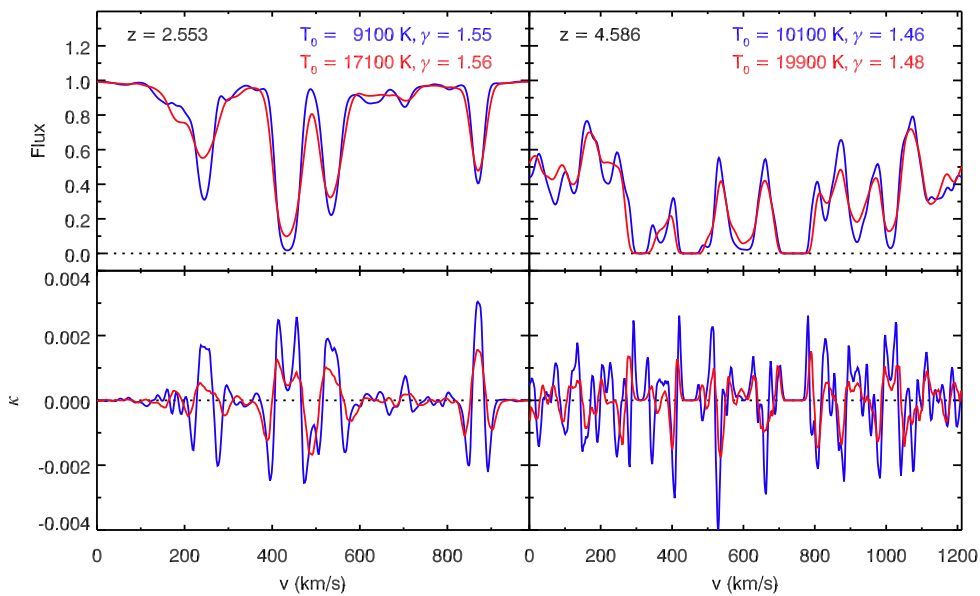}
   \vspace{-0.05in}
   \caption{A demonstration of the curvature statistic.  Top panels: 10\,\hinv\,Mpc sections of simulated \lya\ forest drawn at $z=2.553$ (left) and $z=4.586$ (right).  The sight lines plotted in blue are red drawn from simulations B15 and D15, respectively, which differ by roughly a factor of two in \To\ but have similar \g's.  Bottom panels: The curvature statistic computed from these spectra, as defined in equation~(\ref{eq:curvature}).  At both redshifts, the overall amplitude of $\kappa$ is greater at the lower temperature.}
   \label{fig:curv_example}
   \end{center}
   \end{minipage}
\end{figure*}

Figure~\ref{fig:curv_example} demonstrates the sensitivity of the curvature to the IGM temperature.  We plot simulated spectra at two redshifts, $z = 2.553$ and $z=4.586$, drawn from simulations B15 and D15, for which the temperature at the mean density differs by roughly a factor of two but the slope of the temperature-density relation is the same. At both redshifts, the flux profile drawn from the colder simulation (B15) is significantly more highly peaked, with larger differences between adjacent minima and maxima.  This translates into significantly larger curvature values for the colder simulation, where the peaks in the curvature correspond to the inflection points in the flux.  The overall amount of small-scale structure in a section of forest is measured by computing the mean absolute curvature, \mac\ among pixels within a fixed range range in flux (see Section~\ref{sec:continuum}).  Simulations are then used to convert \mac\ into a temperature, as described in Section~\ref{sec:overdensities}.

\subsection{Sources of Uncertainty}

\subsubsection{Noise}

Several issues must be addressed before the curvature can be used as a reliable probe of the IGM temperature.  The first of these is noise in the flux spectra, which will dominate the curvature measurement if it is calculated directly from even very high-quality data.  Rather than compute the curvature from the raw spectra, therefore, we first fit a smoothly varying b-spline to the flux.  The curvature is then computed from the fit.  We use cubic polynomials for the piece-wise fits, such that the curvature is continuous across the break points.  The b-spline if fit adaptively, with the break points initially separated by 50~\kms.  The fit is then evaluated around each break point, and additional points are added where the fit is poor in a chi-squared sense.  This process is repeated until the fit converges, or until the break point spacing reaches a minimum value (6.7~\kms\ for the HIRES data, 13.6~\kms\ for the MIKE data).

\begin{figure*}
   \centering
   \begin{minipage}{\textwidth}
   \begin{center}
   \includegraphics[width=0.75\textwidth]{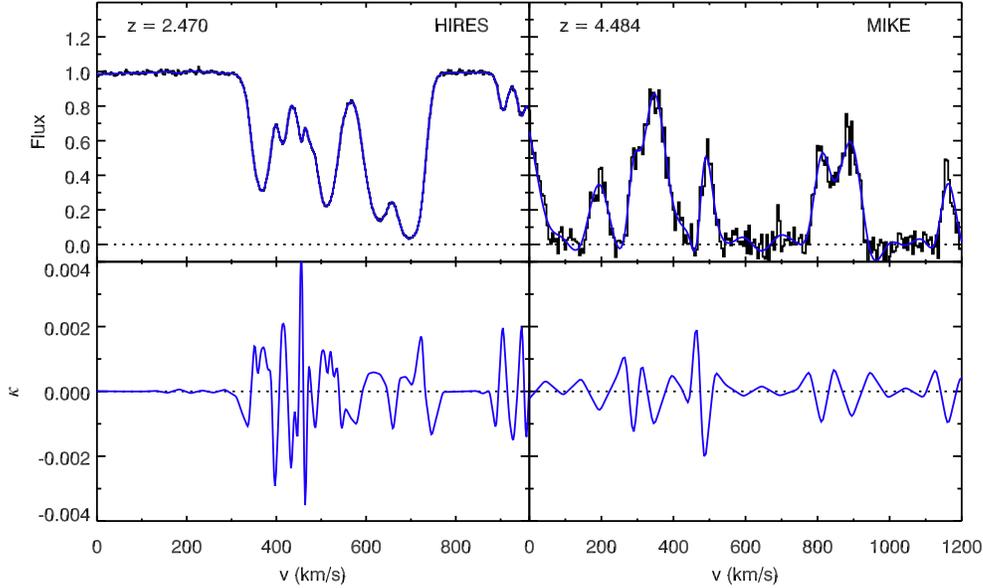}
   \vspace{-0.05in}
   \caption{Top panels: Examples of the \lya\ forest in QSO spectra used in this study, here at $z=2.470$ taken with HIRES (left), and at $z=4.484$ taken with MIKE (right).  Each section spans 10\,\hinv\,Mpc.  The histograms show the normalized spectra.  The continuous lines are b-spline fits.  Bottom panels: The curvature computed from the b-spline fits.}
   \label{fig:bspline_example}
   \end{center}
   \end{minipage}
\end{figure*}

Figure~\ref{fig:bspline_example} shows examples of the b-spline fits to the data.  The fits are naturally more highly constrained for high signal-to-noise data.  In order to compare with the simulations, therefore, we add the same level of the noise to the simulated spectra as we find in the data, and compute the b-splines using the same method.  A significant benefit of our adaptive fitting scheme is that the curvature of the final fit is only moderately sensitive to the amount of noise in the spectrum.  We estimate that our flux error arrays are generally accurate to 10-20\%.  Tests using simulations indicated that uncertainties in the flux error at this level will translate into errors in the temperature measurements that are smaller than our statistical errors.  We have also verified that the measured curvature is not highly sensitive to moderate correlations in the flux errors, and that a constant error is an adequate approximation in cases where the error scales with the flux, as expected for bright sources.  We found that useful curvation informaiton could be extracted from data with errors in the normalized flux as large as $\sim$0.1 per 2.1\,\kms\ bin.  To be conservative, however, we have limited our analysis to regions of the data where the mean error within a 10\,\hinv\,Mpc section is less than 0.06 per 2.1\,\kms.  This was done to maximize the contrast in curvature between simulation runs with different temperatures, which tends to diminish as sharp features become washed out by noise, and to make it easier to identify metal lines (see Section~\ref{sec:metals}).

\subsubsection{Continuum level}\label{sec:continuum}

The curvature as defined by Eq.~(\ref{eq:curvature}) depends linearly on the amplitude of the flux, which in turn will depend on the continuum level.  Unfortunately, the high levels of absorption in the \lya\ forest make it difficult to reliably establish the continuum in the real data, especially at $z \gtrsim 4$, where even the voids may absorb $\gtrsim 10$ per cent of the flux.  This issue has recently been addressed in the context of measuring the mean transmitted flux in the \lya\ forest.  \citet{fg2008} used simulations to estimate their likely error in the continuum placement as a function of redshift, and then applied a correction to their measured mean flux.  This approach will work so long as the optical depth of the voids is consistent.  We wish to allow for changes in the temperature-density relation, however, which can have a strong effect on the opacity of the voids \citep{becker2007,bolton2008}.

We therefore circumvent the continuum issue by ``re-normalizing'' both the real data and the simulations.  For each 10\,\hinv\,Mpc section of forest (the size of our simulation box), we divide the flux by the maximum value of the b-spline fit in that interval.  For example, the spectra in Figure~\ref{fig:bspline_example} at $z=2.470$ and 4.484 would be divided by 1.00 and 0.87, respectively.  The curvature is then computed from the re-normalized fit.  This procedure forces us to sacrifice a small amount of sensitivity.  With the same mean flux level, hotter simulations result in spectra with lower overall curvature  but more absorption at the transmission peaks than colder simulations.  Re-normalizing the flux therefore increases the curvature more in the hotter simulations, decreasing the contrast in curvature between temperatures.  The advantage, however, is that the continuum placement is handled objectively and consistently, removing a potentially significant source of systematic error.  For example, \citet{fg2008} estimate a continuum placement error of $\sim$14 per cent at $z = 4$.  If left uncorrected, this would translate into an error in the curvature of the same amount, which would in turn produce an error in temperature of $\sim$2\,000\,K.

We measure the mean absolute curvature, \mac, of pixels where the re-normalized b-spline fit falls in the range $0.1 \le F^{\rm R10} \le 0.9$, where the ``R10'' superscript denotes that the re-normalization has been performed in 10\,\hinv\,Mpc sections.  The lower bound omits saturated pixels, which do not contain any information on the temperature.  The upper bound is chosen largely to avoid uncertainties in the the mean absolute curvature due to uncertainties in the overall mean flux.  This is particularly a concern at $z \sim 2-3$, where the flux probability distribution function starts to become sharply peaked near $F = 1$.  Because the flux profile tends to be fairly flat near the continuum at these redshifts, the pixels there will tend to have lower than average absolute curvature values.  Increasing the mean flux, even within the best currently-measured uncertainties \citep[e.g.,][]{fg2008}, significantly increases the number of pixels near the continuum, and will therefore systematically decrease \mac.  By not including pixels with $F^{\rm R10} > 0.9$, we avoid most of this potential uncertainty.

\subsubsection{Mean flux}\label{sec:mean_flux}

\begin{figure}
   \begin{center}
   \includegraphics[width=0.45\textwidth]{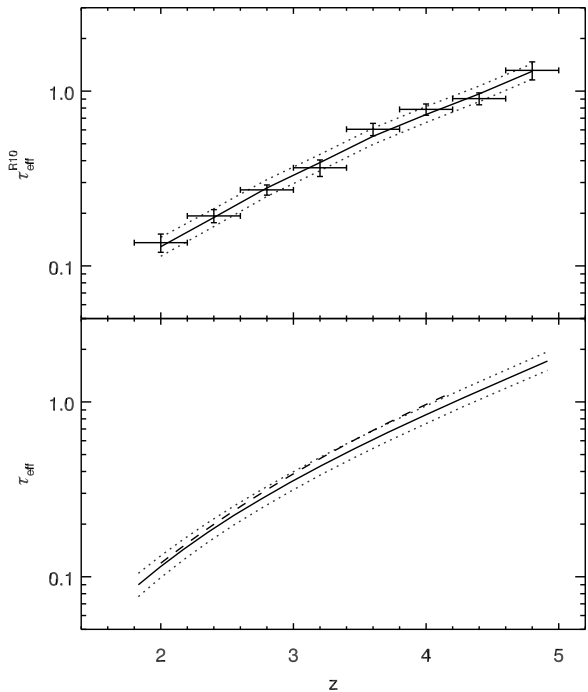}
   \vspace{-0.05in}
   \caption{Top panel:  The effective optical depth measured from our data after the transmitted flux has been re-normalized in 10\,\hinv\,Mpc sections (see Section~\ref{sec:continuum}) and corrected for metal absorption.  Vertical error bars are 2$\sigma$.  The solid line is the fit used to scale the optical depths in our simulations.  The dotted lines show the 2$\sigma$ scatter of the data points about this fit.  Bottom panel: The conventional effective optical depth recovered from the simulations once $\tau^{\rm R10}_{\rm eff}$ has been scaled to match the observations.  The solid and dotted lines correspond to the same data as in the top panel, but here the flux has not been re-normalized.  We use the range bounded by the dotted lines as our 2$\sigma$ uncertainty in $\tau_{\rm eff}$ when calculating uncertainties in the temperature.  For clarity we show only $\tau_{\rm eff}$ from run C15, since the differences for other runs will be much smaller than the 2$\sigma$ uncertainties.  For comparison, $\tau_{\rm eff}$ from \citet{fg2008} is plotted as a dashed line.}
   \label{fig:meanflux}
   \end{center}
\end{figure}

Our choice to exclude pixels near the continuum minimizes the sensitivity of the curvature measurement to the overall mean flux level.  Nevertheless, we have attempted to control for the mean flux in the simulations as carefully as possible.  We use the re-normalized flux spectra to measure a re-normalized effective optical depth, $\tau_{\rm eff}^{\rm R10} = -\ln{\left <F^{\rm R10} \right>}$.  We then adjust the \hi\ ionization rate in the simulations such that $\tau_{\rm eff}^{\rm R10}$ in the artificial spectra matches $\tau_{\rm eff}^{\rm R10}$ in the real data.  This approach has the advantage over matching the true mean flux in that it does not depend on correctly estimating the continuum in the real data.  We also include a correction for metal absorption.  The flux decrement due to metals in the \lya\ forest was estimated by measuring the mean decrement between the \nv~$\lambda1239,1243$ and \siiv~$\lambda1394,1403$ emission lines of the QSOs in our sample.  We found a nearly constant decrement of 3 per cent over $1.8 < z < 4.3$.  The decrement becomes more difficult to measure at higher redshifts as this rest wavelength interval begins to include regions of atmospheric absorption.  However, the uncertainty in the temperature associated with uncertainty in the mean flux becomes small at $z > 3$ (see Section~\ref{sec:Tdelta}).  We therefore adopt a constant 3 per cent correction to the mean \lya\ flux due to metals for all redshifts.

The redshift evolution of $\tau_{\rm eff}^{\rm R10}$ is shown in the top panel of Figure~\ref{fig:meanflux}.  The points show the mean values in the data, corrected for metal absorption, in redshift bins of $\Delta z = 0.4$ along with the 2$\sigma$ uncertainties.  The solid line is the value of $\tau_{\rm eff}^{\rm R10}$ we set in our simulations, while the dotted lines show the 2$\sigma$ scatter of the data points about this fit.  Once the simulations are scaled to produce the correct $\tau_{\rm eff}^{\rm R10}$, the mean flux without re-normalizing can be measured from the artificial spectra.  This is useful for comparing our mean flux constraints to results from the literature.  We plot $\tau_{\rm eff}$ in the bottom panel of Figure~\ref{fig:meanflux}, where $\tau_{\rm eff} = -\ln{\left <F \right>}$.  For clarity, we only show $\tau_{\rm eff}$ for the C15 simulation.  The recovered mean flux will be slightly different for simulations with different temperatures; however, the spread in values is relatively small compared to the uncertainty about the trend shown.  We also plot $\tau_{\rm eff}$ from \citet{fg2008}, which agrees well with our measurements, particularly at $z < 3$, where our method is most sensitive to the mean flux (see Section~\ref{sec:Tdelta}).

\subsubsection{Metal lines}\label{sec:metals}

Metal lines are a potentially serious source of systematic error in any measure of the small-scale structure in the \lya\ forest.  The individual components of metal absorption lines systems tend to be significantly narrower than \lya\ lines \citep[e.g.,][]{wolfe2005}.  If regions of the forest contaminated by metal lines are not properly masked, therefore, the temperature measurements will be significantly biased towards low values.

We identify and mask metal lines using a two-step process.  First, we identify as many absorption systems as possible either from very strong \lya\ absorption (i.e., damped \lya\ systems and Lyman limit systems) or metal lines that fall redward of the \lya\ forest.  The latter may include high-ionization species such as \civ, low-ionization species such as \oi, \siii, \cii, and \feii, or any combination of lines that can be identified as belonging to the same system.  We then mask all regions in the forest that could plausibly contain metal lines at the same redshifts as these systems.

The next step is to search for any remaining unidentified metal lines in the forest.  At high spectral resolution, metal absorption lines tend to stand out from the \lya\ forest by having one or more narrow ($b \lesssim 10-15$~\kms) components, or very sharp edges when saturated.  We therefore mask any lines that appear conspicuously narrow or sharp, as well as any doublets (e.g., \civ, \siiv, \mgii) or other groups of lines that can be identified as belonging to a common system.  

Metal lines are a potentially greater contaminant at lower redshifts, where there are fewer \lya\ lines.  However, this also makes the metal lines easier to identify and remove.   While no metal rejection scheme can be perfect, we expect that we have removed the large majority of metal lines that may significantly contaminate the curvature measurement.  Since we are computing the mean absolute curvature across all pixels, the signal from any remaining unidentified metal lines should be weak in comparison to the signal from the \lya\ forest.  
For example, we have estimated the impact of unidentified \civ\ systems, which generally cannot be identified using other lines redward of the forest.    Random \civ\ systems were added to the simulated spectra at a level consistent with the observed number density and column density distribution over $z \sim 2-4$ \citep{songaila2001}.  Even if we assume that we have missed all systems with $N \le 10^{14}~{\rm cm}^{-2}$, where the upper limit corresponds to a flux decrement near unity, and that these systems are all narrow ($b = 10$~\kms), \mac\ in run C15 at $z \simeq 4.6$ would only increase by $\sim$0.01~dex, which is equivalent to a decrease in \To\ of $\sim$400~K.   At $z \simeq 3.0$, the increase in \mac\ would be $\sim$0.02~dex, corresponding to a decrease in \To\ of $\sim$700~K.  We regard this as a conservative upper limit, since at $z \lesssim 3$ the \lya\ forest is transparent enough that strong doublets can often be identified and masked.  We therefore expect the curvature measurement to be strongly dominated by the genuine \lya\ forest, and that the bias from unidentified metal lines will be smaller than our measurement errors (see Section~\ref{sec:results}).     
 
\subsection{Optimal overdensities}\label{sec:overdensities}

\begin{figure*}
   \centering
   \begin{minipage}{\textwidth}
   \begin{center}
   \includegraphics[width=0.75\textwidth]{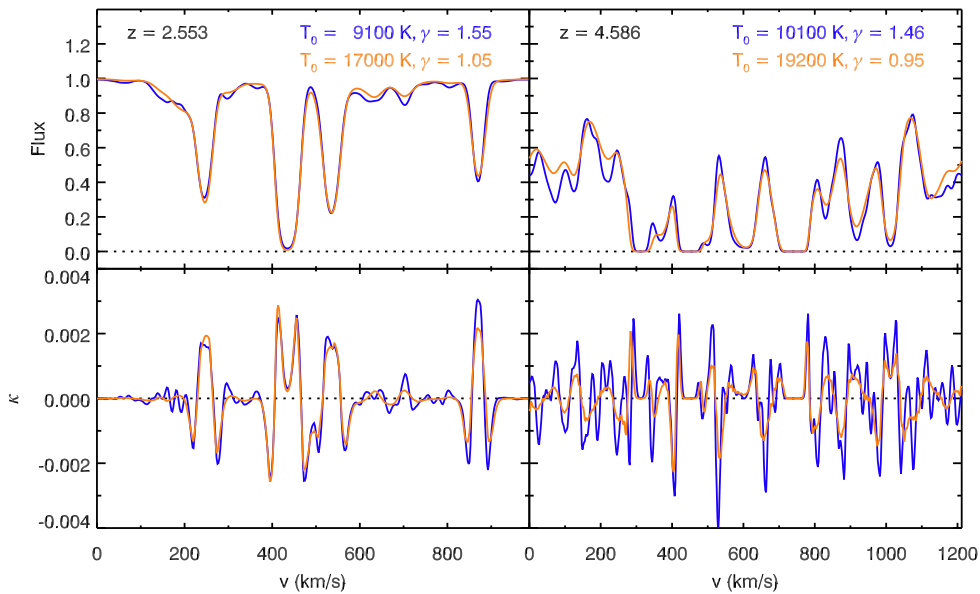}
   \vspace{-0.05in}
   \caption{A demonstration of the degeneracy between \To\ and \g\ in determining the small-scale structure of the \lya\ forest.  The panels are the same as in Figure~\ref{fig:curv_example}, except that run D15 has been replaced with run D10, which has a smaller value of \g.  At $z=2.553$ the \lya\ forest is sensitive to moderate overdensities, where runs B15 (blue lines) and D10 (orange lines) have similar temperature.  Consequently, the \lya\ forests appear very similar and have similar curvatures.  At $z=4.586$, the forest is more sensitive to gas near the mean density.  A change in \g\ therefore does less to alter the curvature.}
   \label{fig:degeneracy}
   \end{center}
   \end{minipage}
\end{figure*}

A fundamental challenge to measuring IGM temperatures is that the temperature will vary with density.  Following reionization the gas in the IGM will be in photoionization equilibrium. At low densities  ($\Delta \lesssim 10$), the heating and cooling of the gas are dominated by the adiabatic contraction and expansion of the filaments and voids due to structure formation, and by the photoionization heating and recombination cooling from  \hi, \hei\ and/or \heii.  The balance between the cooling and heating processes has been shown to produce a relatively tight, gradually steepening power law temperature-density  relation of the form $T(\Delta) = T_{0}\Delta^{\gamma-1}$, where \To\ is the temperature at the mean density \citep{huignedin1997}.  The slope is expected to asymptotically reach $\gamma-1 \approx 0.6$,  with the normalization set by the hardness of the ionizing background.  

Reionization will complicate this picture, however.  The photoheating from reionization will be roughly independent of density.  If reionization were homogeneous and instantaneous, therefore, the temperature-density relation  immediately afterward would be nearly isothermal.  In reality, reionization is likely to be an extended and inhomogeneous process.  During \heii\ reionization, photoheating within growing \heiii\ bubbles around bright QSOs should partially flatten the volume-averaged temperature-density relation as well as introduce significant spatial fluctuations in temperature \citep{gleser2005, furoh2008b, mcquinn2009a, boltonohfur2009a}.  At a minimum, therefore, measurements of the thermal state of the IGM must allow for changes in both the normalization and the slope of the temperature-density relation.  

Even constraining both \To\ and \g\ in a volume-averaged sense is challenging, however, since the dynamic range of the \lya\ forest is limited to densities that produce measurable but non-saturated absorption.  Conventionally, the temperature-density relation is normalized at the mean density ($\Delta = 1$).  From an interpretive standpoint, this is convenient in that a parcel of gas near the mean density will tend to remain near the mean density as the IGM evolves \citep[e.g.,][]{huignedin1997}.  Measuring \To\ as a function of redshift can therefore provide a straightforward indication of how much heating or cooling is occurring for a relatively consistent population of baryons in the IGM.  

The difficulty with measuring \To, however, is that the \lya\ forest does not always trace gas near the mean density.  Ignoring the effects of peculiar velocities and thermal broadening, the \lya\ optical depth at a given overdensity will scale as $\tau(\Delta) \propto H^{-1}(z) \, n_{\rm H\,I} \propto (1+z)^{4.5}\Gamma^{-1}T_0^{-0.7}\Delta^{2-0.7(\gamma-1)}$, where $H(z)$ is the Hubble parameter, $n_{\rm H\,I}$ is the neutral hydrogen density, and $\Gamma$ is the \hi\ photoionization rate \citep[e.g.,][]{weinberg1997, mcdonald2001a}.  The strong redshift dependence makes it extremely challenging to probe a constant range of overdensities over all redshifts.  In practice, the \lya\ forest will tend to be sensitive to gas that produces optical depths near unity, which occurs near the mean density at $z \sim 4-5$, and at mild overdensities at $z \sim 2-3$.  At lower redshifts, therefore, the measurement of \To\ will become increasingly degenerate with \g, since most of the signal will be coming from densities above the mean.  

The degeneracy of \To\ and \g\ in setting the amount of small-scale structure in the \lya\ forest is demonstrated in Figure~\ref{fig:degeneracy}.  Here we show the same simulated sections of \lya\ forest as in Figure~\ref{fig:curv_example}, but with simulation D15 replaced by D10, which has a nearly isothermal temperature-density relation ($\gamma \sim 1$).  The gas at $\Delta > 1$ is colder in this simulation than it would be in the case of $\gamma \sim 1.5$.  Consequently, at $z = 2.553$ the \lya\ forest in B15 and D10 appears nearly identical, even though \To\ differs by roughly a factor of two.  The similarity is also seen in the curvature values.  At $z = 4.586$ the fluctuations in the forest are still significantly smoothed out in run D10 as compared to run B15, since the forest at this redshift is more sensitive to gas near the mean density than at lower redshifts.

In principle one can take advantage of the fact that lower-column density absorbers tend to arise from lower-density gas in order to constrain both \To\ and \g\ \citep[e.g.,][]{schaye2000,theuns2000b}.  In practice, however, this has proven to be extremely challenging, and the error bars on both quantities have remained large.  At $z > 4$ the forest becomes increasingly difficult to separate into discrete absorbers, posing a further obstacle from specifying the full temperature-density relation from the \lya\ forest alone.

\begin{figure}
   \centering
   \begin{center}
   \includegraphics[width=0.45\textwidth]{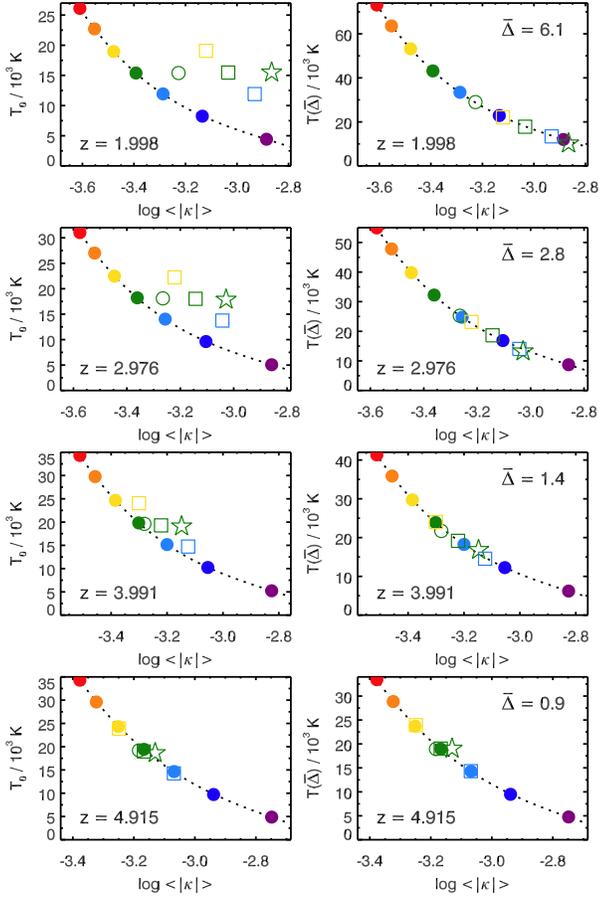}
   \vspace{-0.05in}
   \caption{Left panels: \To\ as a function of \logmac\ for our main simulation runs.  Colors correspond to different values of \To, as in Figure~\ref{fig:sim_grid}.  Symbols denote different values of \g\ -- {\it solid circles}: $\gamma \sim 1.5$, {\it open circle}: $\gamma \sim 1.3$, {\it boxes}: $\gamma \sim 1.0$, {\it star}: $\gamma \sim 0.7$.  The correspondence between the curvature and \To\ becomes increasingly confused at low redshifts, where not knowing $\gamma$ can lead to a very wrong estimate of \To.  Right panels: The same, except that the temperatures at each redshift are measured at an optimal overdensity that has been chosen such that \Tdelta\ is a nearly one-to-one function of \logmac, regardless of $\gamma$.}
   \label{fig:T_delta_vs_curvature}
   \end{center}
\end{figure}

Rather than attempt to measure both the normalization and the slope of the temperature-density relation, therefore, we use the curvature to measure the temperature at an {\it optimal overdensity} probed by the \lya\ forest.  The optimal overdensity, $\bar{\Delta}$, which will change with redshift, is empirically determined to be the overdensity at which \Tdelta\ is a simple function of \mac, regardless of \g.  Our approach is illustrated in Figure~\ref{fig:T_delta_vs_curvature}.  In the left-hand panels we plot \To\ as a function of \logmac\ for the various simulation runs.  The simulations with $\gamma \sim 1.5$ form a well-defined relationship between the curvature and \To.  However, runs with different \g's can be well off of this relationship, particularly at lower redshifts.  For a given \logmac\ at $z \simeq 2$, the difference in \To\ between $\gamma \sim 1.5$ and $\gamma \sim 1.0$ can be $\gtrsim 10\,000$~K.  This is clearly undesirable when \g\ is poorly constrained.  In the right-hand panels of Figure~\ref{fig:T_delta_vs_curvature}, in contrast, we plot the temperature at an optimal overdensity that has been empirically chosen to minimize the variations in the relationship between \Tdelta\ and \logmac.  In this case, \Tdelta\ is a nearly one-to-one function of \logmac\ regardless of \g.
These values can be interpreted as characteristic overdensities probed the \lya\ forest, though in practice the curvature measurement will be an average over a range of densities. The important point is that we can we can identify a value of $\bar{\Delta}$ where \Tdelta\ is a simple function of the curvature, which allows us to reliably measure the temperature at a density probed by the \lya\ forest without knowing the full form of the temperature-density relation.  \citet{mcdonald2001b} used a similar approach in their line-fitting analysis, fitting $T$ and \g\ at an overdensity where the errors were uncorrelated.

\begin{figure}
   \centering
   \begin{center}
   \includegraphics[width=0.45\textwidth]{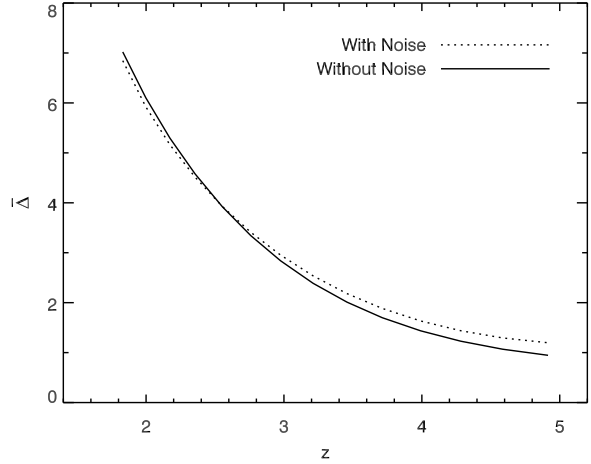}
   \vspace{-0.05in}
   \caption{The optimal overdensity at which \Tdelta\ is a nearly one-to-one function of the curvature.  The solid line shows $\bar{\Delta}(z)$ for noise-free spectra, while the dotted line shows $\bar{\Delta}(z)$ for spectra with noise properties similar to the real data.  The noise amplitude increases with redshift, where it creates a mild bias in the curvature measurement towards higher overdensities.}
   \label{fig:overdensity}
   \end{center}
\end{figure}

The optimal overdensities for the curvature analysis will depend somewhat on the details of the method and on the quality of the data.  We have selected our overdensities after computing \mac\ from the simulations using the method outlined in Section~\ref{sec:method}.  The results are shown in Figure~\ref{fig:overdensity}.  We computed the optimal overdensities once using noise-free spectra and again after noise was added to the simulations at levels consistent with the real data.  The results are very similar, although adding noise tends to produce a slight bias towards higher overdensities at $z > 3$.  In our analysis we use the optimal overdensities computed with noise, and determine the temperatures based on the curvature values of the simulations with $\gamma \sim 1.5$.  Deviations from the \Tdelta-\mac\ relationship for other values of \g\ are typically less than 200-400~K, and are always less than our final $1\sigma$ uncertainty in \Tdelta.  Finally, we note that even at redshifts where the optimal overdensity is several times the mean density the temperature measurement will be dominated by gas that is photoheated, although shock heating begins to make a minor contribution at $z \lesssim 2$.

\subsection{Summary of Method}\label{sec:method_summary}

Our method for measuring temperatures from the curvature of the \lya\ forest is thus summarized as follows:

\begin{enumerate}

\item{A b-spline is fit to the data.  We use an initial break point spacing of 50~\kms, adding more break points as needed to achieve a good fit.  The minimum break point spacing is set by the instrumental FWHM.}

\item{The spectra are divided into 10\,\hinv\,Mpc sections, and the flux is re-normalized by the maximum value of the b-spline fit in each section.}

\item{The curvature, $\kappa$, is computed from the re-normalized b-spline fit.}

\item{The path length-weighted mean absolute curvature, \mac, is computed from all unmasked pixels where the re-normalized fit falls in the range $0.1 \le F^{\rm R10} \le 0.9$.  We group the data in redshift bins of $\Delta z = 0.4$.}

\item{For each 10\,\hinv\,Mpc section of real data, the simulated spectra from each simulation run are smoothed to the same instrumental resolution and rebinned to the same pixel size.  The \mac\ values are then computed from the synthetic spectra using the procedure described above after noise has been added and b-spline fits made.  To save computational time, a grid of noise values are used ($r.m.s. = [0.005,0.01,0.02,0.04,0.08]$) and the values of \mac\ are interpolated between noise levels and simulation redshifts to match the data.  A path length-weighted \mac\ is computed for each simulation using the same weighting as was used for the real data in each redshift bin.}

\item{\Tdelta\ is determined by interpolating the \Tdelta\ verses \logmac\ relationship in the simulations to the the value of \logmac\ measured from the data.  The overdensity, $\bar{\Delta}$, is the optimal overdensity plotted in Figure~\ref{fig:overdensity}.}

\end{enumerate}

\subsection{Impact of Thermal History}\label{sec:thermal_history}

The small-scale structure of the \lya\ forest, as measured by the curvature or any other method, will depend on the thermal broadening of the absorption features as well as on the Hubble broadening and any turbulent broadening.  Heating the gas will increase not only the thermal broadening but also the characteristic physical size of absorbers, an effect known as Jeans smoothing.  The result is an additional smoothing of the \lya\ forest due to greater Hubble broadening across individual absorbers.  In principle, therefore, the small-scale structure of the \lya\ forest will depend both on the instantaneous temperature of the gas and on its integrated thermal history \citep[e.g.,][]{pawlik2009}.

\begin{figure}
   \begin{center}
   \includegraphics[width=0.45\textwidth]{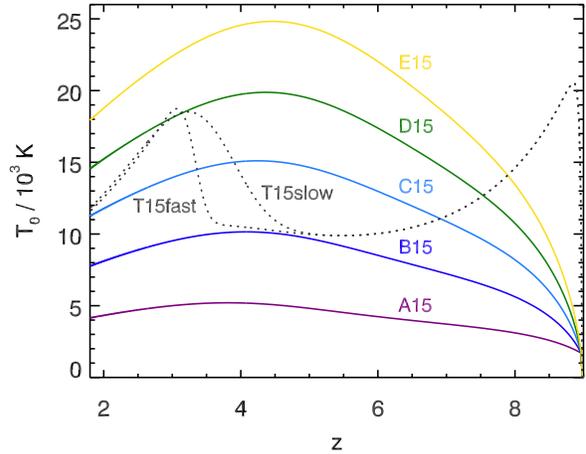}
   \vspace{-0.05in}
   \caption{Thermal histories of a subset of simulations used in this work.  The solid lines are for runs A15-E15, with colors corresponding to Figure~\ref{fig:sim_grid}.  The gas temperature in these runs is set by photoheating in ionization equilibrium, but does not include an initial temperature boost due to hydrogen reionization.  The dotted lines shows the temperature at the mean density for runs T15slow and T15fast, which include heating meant to mimic an instantaneous hydrogen reionization at $z = 9$, followed by either an extended (T15slow) or rapid (T15fast) \heii\ reionization completing near $z \sim 3$.  We apply our curvature method to artificial data drawn from these two runs in order to test the impact of the thermal history on measurements of the instantaneous temperature (see Section~\ref{sec:thermal_history}).}
   \label{fig:histories}
   \end{center}
\end{figure}

\begin{figure*}
   \begin{center}
   \includegraphics[width=0.85\textwidth]{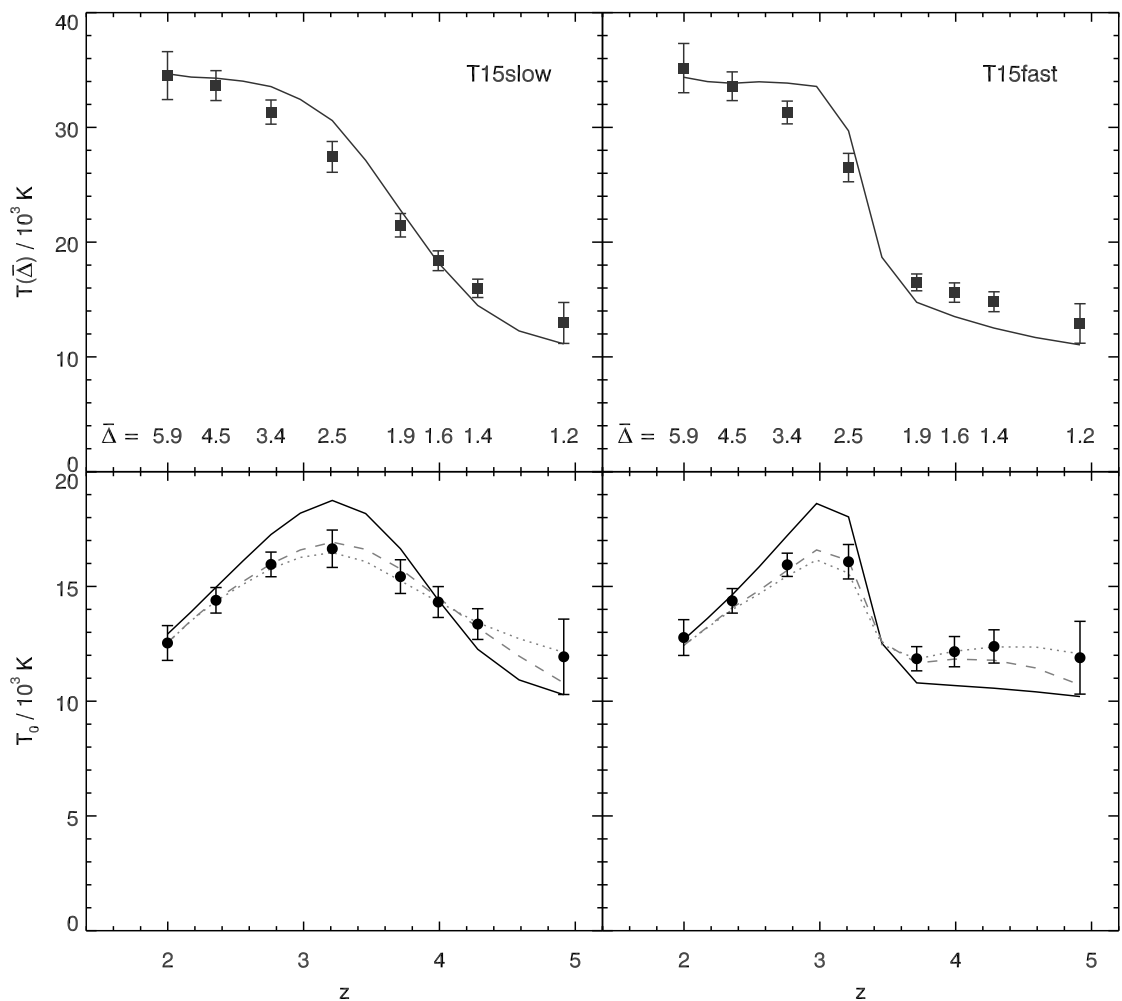}
   \vspace{-0.05in}
   \caption{Temperatures recovered from the test runs using our method.  Left panels are for run T15slow, right panels for T15fast.  Top panels: Temperature at the optimal overdensity.  The overdensities at which \Tdelta\ is measured are printed along the bottom of each plot.  Bottom panesl: Temperature at the mean density inferred for the fiducial $\gamma \sim 1.5$, which was used in both test runs.  In all panels, points with error bars (2$\sigma$) give the values we measure using artificial spectra with resolution, noise properties, and sample size similar to the real data.  The solid lines give the actual values measured directly from the simulation.  The dashed and dotted lines in the bottom panels give the values of \To\ that would be recovered from noise-free HIRES and MIKE data, respectively.  Differences in the thermal history of the test runs and the comparison simulations (cf. Figure~\ref{fig:histories}) cause the increase in \To\ to be underestimated by $\sim$4\,000\,K.}
   \label{fig:test_run}
   \end{center}
\end{figure*}

In order to test the sensitivity of our method to the thermal history of the IGM, we applied our analysis to a set of spectra drawn from simulations whose thermal histories included simple scenarios for hydrogen and helium reionization.  The test runs, T15slow and T15fast, include heating meant to mimic \hi\ reionization at $z = 9$, and either an extended \heii\ reionization spanning $3 < z < 5$ (T15slow) or a rapid \heii\ reionization over $3 < z < 3.5$ (T15fast).  Current estimates of the ionizing emissivity from QSOs suggest that \heii\ reionization should be a patchy and extended process, producing a gradual evolution in the volume-averaged temperature \citep[e.g.,][]{boltonohfur2009a,mcquinn2009a}.  Nevertheless, we include the T15fast model in order to demonstrate that our method is sensitive to rapid changes in temperature, despite the fact that the temperatures of the comparison models evolve slowly.  For the sake of simplicity, the slope of the temperature-density relation is held to the fiducial value ($\gamma \sim 1.5$) in these runs.  The temperature histories of the test runs are compared to runs A15-E15 in Figure~\ref{fig:histories}.  Artificial data were generated using similar redshifts, noise levels, instrumental resolutions, and sample sizes as the real data.  The curvature was then measured using the same procedure described in Section~\ref{sec:method}, with the exception that no metal masking was performed.

The temperatures recovered from the test simulations are plotted in Figure~\ref{fig:test_run}.  The top panels shows the temperature at the optimal overdensity, \Tdelta, with the value of $\bar{\Delta}$ at each redshift printed along the bottom of each plot.  The bottom panels shows \To, which we have computed using the values of \g\ measured in the simulations.  In all panels the recovered temperatures are plotted using filled symbols, while the actual temperature measured directly from the simulations is shown with a solid line.  We also show the values of \To\ that would be recovered from noise-free spectra.  The results at HIRES (6.7~\kms) and MIKE (13.6~\kms) resolution are plotted using dashed and dotted lines, respectively, where the smoothing has been applied to both the test simulations and the comparison runs.    In general, the curvature method recovers the true values very well.  Notably, a gradual increase in \To\ from $z \simeq 5$ to 3 is correctly recovered for T15slow, while a sudden jump in \To\ at $ z \sim 3.5$ is recovered for T15fast.   In both cases, however, the measured  temperatures show some deviation from the true values.  At $z \sim 5$, \Tdelta\ is overestimated by $\sim$2\,000\,K, while at $z \sim 3$, \Tdelta\ is underestimated by $\sim$3\,000\,K for both T15slow and T15fast.  The corresponding net rise in \To\ is underestimated by $\sim$4\,000\,K.  

The errors in the recovered temperatures can be understood in terms of the differences in thermal history between the test runs and runs A15-G15, which were used to calibrate the conversion from \mac\ to \Tdelta.  At $z \sim 5$, the T15 runs (which are identical at $z > 5$) and run B15 have similar values of \To.  Since the T15 runs included a heat injection at $z \sim 9$, however, the Jean smoothing in these runs will be substantially higher than in run B15.  This should cause the T15 runs to have a comparatively lower overall curvature at $z \sim 5$, leading to a higher estimate of the instantaneous temperature.  We note, however, that at $z \sim 5$ the impact of the thermal history on the measured temperature depends partially on the properties of the data.  The recovered value of \To\ in the highest-redshift bin is very near the true value (solid line) for noise-free HIRES data (dashed line), but increases at MIKE resolution (dotted line) or when noise is added (filled point, for which 90 per cent of the spectra are at HIRES resolution).  These effects relate to properties of the \lya\ forest that are sensitive to the amount of Jeans smoothing.  For a given overall mean flux, a lower amount of Jeans smoothing at $z \sim 5$ will tend to produce sharper and more distinct transmission peaks, which will dominate the curvature signal.  With increased Jeans smoothing, more of the high-curvature regions for the same \To\ will instead be small undulations on top of broader transmission regions.  Such undulations are more easily obscured than distinct peaks by smoothing or adding noise.  The curvature at $z \sim 5$ in the T15 runs is therefore lower than in run B15 when measured from present-quality spectra, and so the temperature in the test runs is over-estimated.

The impact of thermal history is more straightforward at $z \sim 3$.  Here, changes in the temperature measurement due to noise and lowered resolution are minimal.   The T15 runs and run D15 have similar values of \To, but because the gas in the T15 runs was colder than in run D15 at higher redshifts it has experienced less Jeans smoothing.  The curvature in the T15 runs at $z \sim 3$ is therefore higher by comparison, which leads to an underestimate of the temperature.

Although the instantaneous temperature is exactly recovered at certain redshifts, this does necessarily mean that the gas has returned to hydrostatic equilibrium at these points.  The \lya\ forest absorbers at a given density will tend to have physical sizes on order of the local Jeans length, $L_{\rm J} \sim c_{\rm s}  \, t_{\rm dyn}$, where $c_{\rm s}$ is the sound speed and $t_{\rm dyn}$ is the dynamical timescale \citep{schaye2001}.  Following a heating event, therefore, local hydrostatic equilibrium will only be restored on the dynamical timescale, which at the mean density is on the order of a Hubble time: $t_{\rm dyn} \equiv 1 / \sqrt{G \rho} \sim t_{\rm H} \, \Delta^{-1}$.  Even at $z = 2-3$, where the forest is probing mildly overdense regions, the thermal response time for typical absorbers will be significantly longer than the timescales over which hydrogen and helium reionization occur.  In that case, the small-scale structure of the \lya\ forest is {\it always} sensitive to the thermal history of the gas.  The redshifts where we exactly recover the temperatures in the test simulations are simply those where the integrated heat input in the T15 runs has produced a similar amount of Jeans smoothing as in the comparison run with the same instantaneous temperature.  We will revisit the implications of this test for our temperature measurements from the real data in the next section.

\section{Results}\label{sec:results}

\subsection{Curvature}

\begin{figure}
   \begin{center}
   \includegraphics[width=0.45\textwidth]{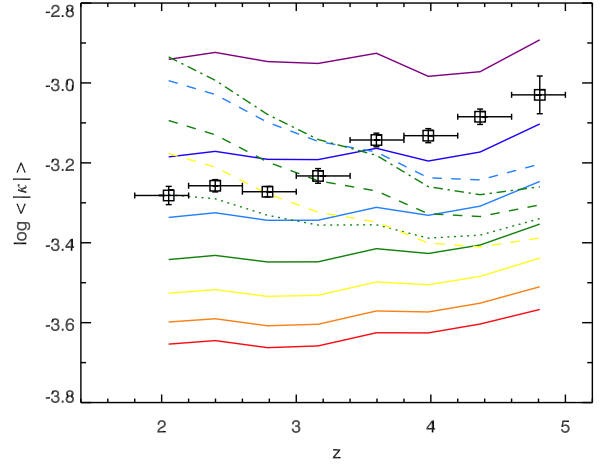}
   \vspace{-0.05in}
   \caption{Raw curvature measurements.  Boxes with error bars are for the real data.  Vertical error bars are 2$\sigma$, while horizontal error bars show the redshift range spanned by each bin.  Lines show the curvature from the simulations using the same mix of resolution and noise properties present in the data at each redshift.  Variations in the resolution and noise cause the curvature in the simulations to be non-smooth functions of redshift.  Line styles and colors correspond to Figure~\ref{fig:sim_grid}.}
   \label{fig:curv_measurements}
   \end{center}
\end{figure}

\begin{figure*}
   \centering
   \begin{minipage}{\textwidth}
   \begin{center}
   \includegraphics[width=0.75\textwidth]{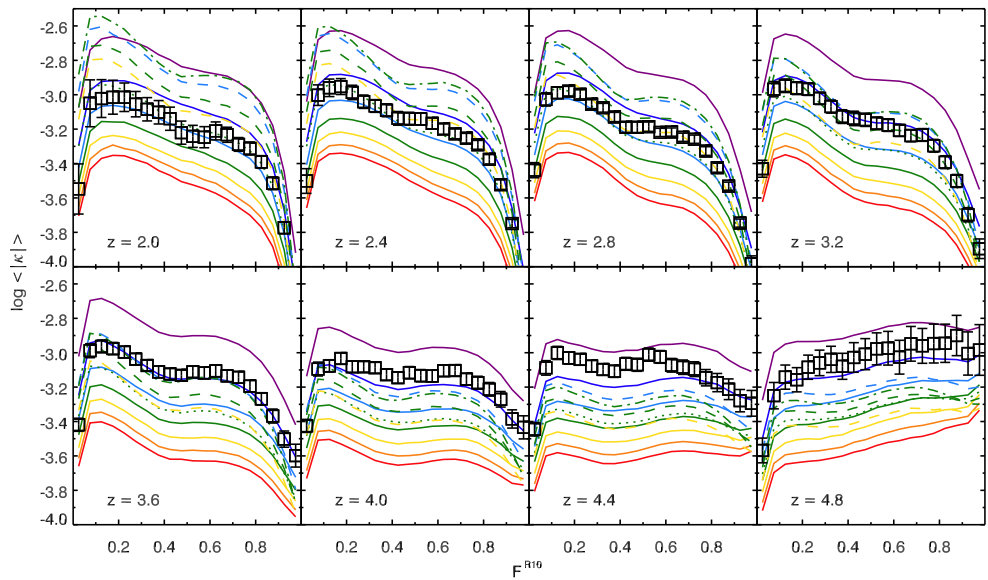}
   \vspace{-0.05in}
   \caption{Mean absolute curvature values as a function of re-normalized flux.  Boxes with error bars are for the real data.  The error bars are $2\sigma$, though adjacent flux bins are highly correlated.  Lines give values from the simulations using the same mix of resolution and noise properties present in the data at each redshift.  In general, the observed dependence of \logmac\ on flux is well reproduced by the simulations.  Line styles and colors correspond to Figure~\ref{fig:sim_grid}.}
   \label{fig:curv_vs_flux}
   \end{center}
   \end{minipage}
\end{figure*}

We first present our results for the raw curvature measurements.  In Figure~\ref{fig:curv_measurements} we plot \logmac\ as a function of redshift.  The data points are plotted with $2\sigma$ vertical error bars in redshift bins of $\Delta z = 0.4$.  The uncertainties were estimated from the simulations using sets of artificial data with similar redshifts, noise levels, spectral resolution, and sample size as the real data. In redshift bins containing large numbers ($\mathcal{N} > 100 $) of data points, these agreed very well with bootstrap errors generated directly from the data.  Values of \logmac\ from the simulations are also plotted.  These were also calculated using a similar mix of redshifts, noise levels, and spectral resolutions as the real data but from the full set of simulated sight lines.  Variations in the properties of the data introduce deviations from the otherwise smooth evolution of the curvature with redshift (cf. Figure~\ref{fig:sim_grid}).  The uncertainties in our measured values of \logmac\ are typically $\lesssim 20$\% of the separation between adjacent simulations with $\gamma \sim 1.5$, though the error is somewhat larger at $z = 4.8$, where the sample size is significantly smaller.

As a consistency check, we have also computed \logmac\ as a function of the re-normalized flux level.  The results for each redshift bin are shown in Figure~\ref{fig:curv_vs_flux}.  Uncertainties in the data points were again estimated from the simulations.  The errors in neighboring flux bins will be highly correlated, but we note that the shape of \logmac\ verses \Fr\ in the simulations is in good agreement  with that of the real data.  While we use the overall mean absolute curvature to measure the temperatures, this agreement provides assurance that the simulated spectra exhibit realistic curvature properties.  We note that the shape of the curvature as a function of flux is moderately dependent on \g.  This is especially true at lower redshifts, where smaller values of \g\ produce higher peaks near $F^{\rm R10} = 0.1$ and a more pronounced inflection near $F^{\rm R10} = 0.4$.  Setting constraints on \g\ is presently somewhat difficult due to  the correlations in the measurement errors and the fact that we have only run a small number of simulations with $\gamma < 1.5$.  We therefore leave this for a future work.

\subsection{Temperature at the optimal overdensity}\label{sec:Tdelta}

The primary result from this study is the evolution of the IGM temperature at the optimal overdensity probed by the \lya\ forest, which we infer directly from the curvature measurements.  Our values for \Tdelta\ are given in Table~\ref{tab:temp}, and plotted in Figure~\ref{fig:T_delta}.  For each redshift bin we print the optimal overdensity at which \Tdelta\ is measured along the bottom of the plot.  The $2\sigma$ errors are again estimated using the simulations, and include the errors related to uncertainties in \taueff\footnote{These tend to be small since we are restricting our analysis to the center of the flux distribution.  The mean absolute curvature does vary with flux, as shown in Figure~\ref{fig:curv_vs_flux}.  However, the shape of the flux PDF in the range $0.1 \le F \le 0.9$ does not change quickly as \taueff\ changes, and the flux-weighted \mac\ therefore remains relatively constant.  The $2\sigma$ error in \Tdelta\ related to uncertainty in \taueff\ is less than 400\,K at $z \ge 2.8$, increasing to $\sim$1300\,K at $z = 2.0$.}.  The median error is 8 per cent, although this does not include uncertainties related to the thermal history.  

A strong trend of increasing \Tdelta\ with decreasing redshift is evident.  Unless the temperature-density relation is flat or inverted over this entire interval, which appears unlikely, much of this rise in temperature is probably  due to the fact that the overdensity probed by the  \lya\  forest data with our measurements increases with redshift, from $\bar{\Delta} = 1.2$ at $z = 4.8$ to $\bar{\Delta} = 5.7$ at $z=2.0$. The lower density gas is adiabatically cooled as the voids expand, whereas the moderate density gas is heated as the filaments collapse.  For the relatively steep temperature-density relation  expected following hydrogen reionization, the higher density gas probed at lower redshifts is therefore expected to be hotter.  Interpreting the \Tdelta\ results in the context of \heii\ reionization is thus not straightforward, and requires a comparison  to simulations of \heii\ reionization in which the full temperature-density relation is modeled.

\begin{table*}
   \caption{Temperature results.  Columns give the redshift interval, the path length-weighted mean redshift, the optimal overdensity, the total number of 10\,\hinv\,Mpc sections included, the temperature at the optimal overdensity, the fiducial value of \g, the temperature at the mean density for the fiducial \g, and the temperature at the mean density for $\gamma = 1.3$.  All errors are 2$\sigma$.}
   \label{tab:temp}
   \begin{minipage}{\textwidth}
   \begin{center}
   \begin{tabular*}{\textwidth}{@{\extracolsep{\fill}}cccccccc}
   \hline
    z  &  $\langle z \rangle$  &  $\bar{\Delta}$  &  $\mathcal{N}$  &  $T(\bar{\Delta})$  &  $\gamma^{\rm fid}$  &  $T_{0}^{\rm fid}$  &  $T_{0}^{\gamma = 1.3}$ \\ 
   \hline
      $1.8 - 2.2$  &  2.05  &  5.69  &  202  &  $ 28670 \pm 2200  $  &  1.57  &  $ 10680 \pm  820\ $  &  $ 17020 \pm 1300  $  \\
      $2.2 - 2.6$  &  2.40  &  4.39  &  313  &  $ 25430 \pm 1320  $  &  1.56  &  $ 11080 \pm  580\ $  &  $ 16320 \pm  840\ $  \\
      $2.6 - 3.0$  &  2.79  &  3.35  &  340  &  $ 22630 \pm  960\ $  &  1.55  &  $ 11640 \pm  500\ $  &  $ 15740 \pm  660\ $  \\
      $3.0 - 3.4$  &  3.16  &  2.62  &  120  &  $ 18460 \pm 1180  $  &  1.54  &  $ 10980 \pm  700\ $  &  $ 13820 \pm  880\ $  \\
      $3.4 - 3.8$  &  3.60  &  2.02  &  114  &  $ 13870 \pm  960\ $  &  1.52  &  $\ 9590 \pm  680\ $  &  $ 11230 \pm  780\ $  \\
      $3.8 - 4.2$  &  3.98  &  1.64  &  100  &  $ 11160 \pm 1000  $  &  1.51  &  $\ 8670 \pm  780\ $  &  $\ 9610 \pm  860\ $  \\
      $4.2 - 4.6$  &  4.37  &  1.40  &   92  &  $\ 9330 \pm 1040  $  &  1.49  &  $\ 7920 \pm  880\ $  &  $\ 8440 \pm  940\ $  \\
      $4.6 - 5.0$  &  4.81  &  1.23  &   22  &  $\ 8930 \pm 2020  $  &  1.46  &  $\ 8120 \pm 1840  $  &  $\ 8390 \pm 1900  $  \\
   \hline
   \end{tabular*}
   \end{center}
   \end{minipage}
\end{table*}

\subsection{Temperature at the mean density}

\begin{figure}
   \begin{center}
   \includegraphics[width=0.45\textwidth]{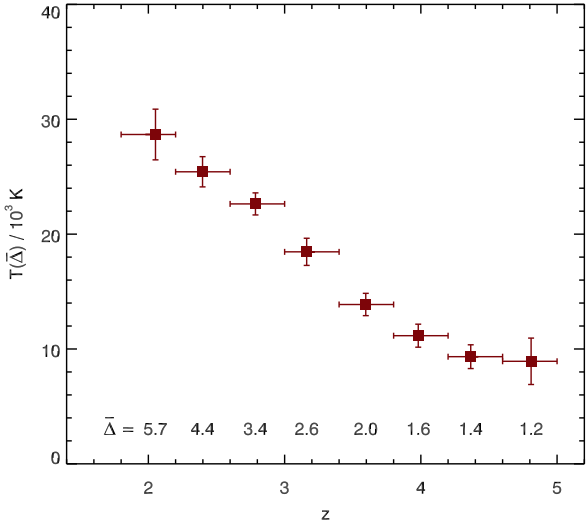}
   \vspace{-0.05in}
   \caption{IGM temperature at the optimal overdensity as a function of redshift.  These are the primary measurements obtained by this study.  The overdensity for each redshift bin is printed along the bottom of the plot.  Vertical error bars are 2$\sigma$, and include statistical errors estimated from the simulations along with errors associated with the uncertainty in the mean flux.  Horizontal error bars indicate the redshift range spanned by each bin.}
   \label{fig:T_delta}
   \end{center}
\end{figure}

\begin{figure}
   \begin{center}
   \includegraphics[width=0.45\textwidth]{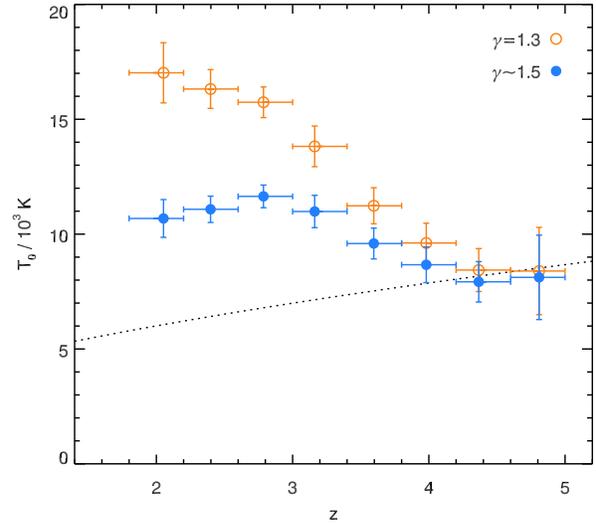}
   \vspace{-0.05in}
   \caption{Temperature at the mean density inferred from our \Tdelta\ measurements under different assumptions for \g.  Filled circles are for the fiducial values of \g\ listed in Table~\ref{tab:temp}, which are typically near 1.5.  This is the maximum slope expected in photoionization equilibrium, and hence the filled circles represent minimum values for \To.  Open circles give \To\ for $\gamma = 1.3$.  \To\ for an isothermal IGM ($\gamma = 1.0$) would be given by the \Tdelta\ values in Figure~\ref{fig:T_delta}.  Note that \To\ at $z > 4$ is largely insensitive to \g, since the optimal overdensity at which we measure \Tdelta\ is very close to the mean density (cf. Figure~\ref{fig:T_delta}).  The dotted line is a thermal asymptote of the form $T_{0} \propto (1+z)^{0.53}$ that has been scaled to approximately match our measurements at $z \ge 4.4$.  It shows the expected temperature evolution in the absence of \heii\ reionization \citep{huihaiman2003}.}
   \label{fig:T0}
   \end{center}
\end{figure}

Translating our \Tdelta\ measurements into temperatures at the mean density requires knowing the slope of the temperature-density relation.  While we have not attempted to fit the full relation, we can infer what \To\ should be for reasonable values of \g.  We plot \To\ in Figure~\ref{fig:T0} under two different assumptions.  In the first case, we use the values of \g\ measured in our fiducial simulations (A15-G15), where \g\ increases from $\gamma \approx 1.4$ at $z = 5$ to $\gamma \approx 1.6$ at $z = 2$.  The exact values for \g\ are given by the solid lines in the middle panel of Figure~\ref{fig:sim_grid} and in Table~\ref{tab:temp}.  In this case, shown by the filled circles, \To\ is consistent with $\sim$8\,000\,K at $z \ge 4.4$, increasing to $\sim$12\,000\,K at $z = 2.8$, and then decreasing to $\sim$11\,000\,K at $= 2$.  This scenario represents the minimum \To\ case, in which there is no additional  heating in the voids that would tend to flatten the temperature-density relation.  Even in this case, an increase in \To\ between $z = 4.4$ and $z = 2.8$ is detected at the 7$\sigma$ level.

In the second case, we assume a constant $\gamma = 1.3$.   A value
of  $\gamma$ in this range is suggested by the numerical simulations
of \citet{mcquinn2009a}. It corresponds to a  mild flattening of 
the temperature-density relation as expected during an extended \heii\
reionization process.  \To\ is largely insensitive to \g\ at $z > 4$,
where the optimal overdensity at which we measure \Tdelta\ is already
close to the mean density.  At lower redshifts, however, \To\ is
substantially higher for $\gamma = 1.3$, up to $\sim$17\,000\,K at $z
= 2.0$.  Following the end of \heii\ reionization, \g\ is expected to
return to its asymptotic value ($\gamma \sim 1.5$), and so our
measurements are consistent with a decline in \To\ at $z < 2.8$ even
if the temperature-density relation is flattened at higher redshifts.
We emphasize that the \g\ relevant to our study is the
globally-averaged value, even though there may be significant scatter
in the temperature-density relation.  Within growing 
\heiii\ bubbles, dense regions should ionize first and have time to
cool  before the voids become fully ionized \citep{furoh2008b}.  
Radiative transfer effects may also create additional complexity in
the temperature-density relation \citep{boltonohfur2009a}.  Getting
the temperature-density relation to flatten across the entire IGM
simultaneously, however, would require a brief and
spatially-coordinated \heii\ reionization, which appears 
not to be favored on theoretical grounds
\citep[e.g.,][]{mcquinn2009a}.  Nevertheless, if the IGM is isothermal
then \To\ would be given by the values of \Tdelta\ in Figure~\ref{fig:T_delta}.  

\begin{figure*}
   \centering
   \begin{minipage}{\textwidth}
   \begin{center}
   \includegraphics[width=0.85\textwidth]{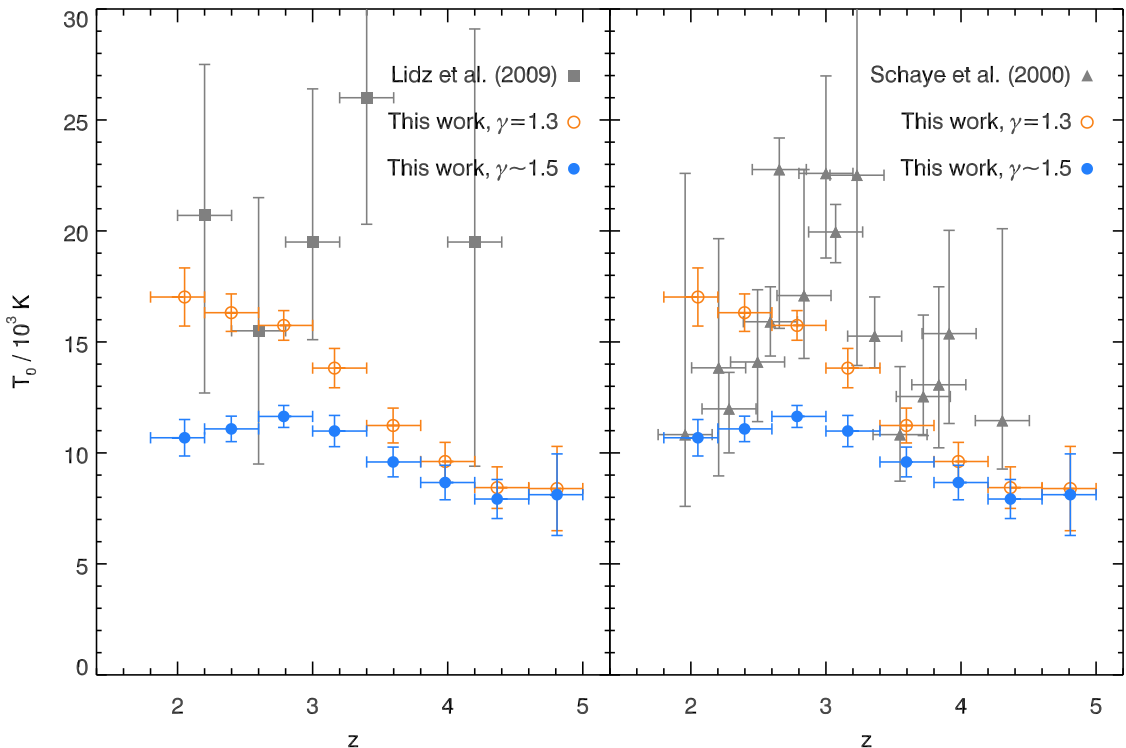}
   \vspace{-0.05in}
   \caption{Comparison of our results with selected results from the literature.  The left-hand panel shows our \To\ points for $\gamma \sim 1.5$ and $\gamma = 1.3$ along with \To\ from the wavelet analysis done by \citet{lidz2009}.  All vertical error bars in this panel are 2$\sigma$.  The right-hand panel shows our results along with those from the Voigt-profile analysis of \citet{schaye2000}.  Vertical error bars for the Schaye et al. results are 1$\sigma$.  See text for discussion.}
   \label{fig:T0_with_lit}
   \end{center}
   \end{minipage}
\end{figure*}

Our test of the impact of thermal history in Section~\ref{sec:thermal_history} suggests that we may have somewhat underestimated the increase in \To\ from $z > 4$ to $z \sim 3$.  We emphasize, however, that since we are comparing the data to simulations with flat thermal histories, the position of the peak in $T_0(z)$ for the $\gamma \sim 1.5$ case in Figure~\ref{fig:T_delta} is likely to be correct.  Future measurements of the spatial coherence of the absorbers using pairs of QSO sight lines should allow us to disentangle the Jeans smoothing from the instantaneous temperature \citep{peeples2010b}.  For now, the detection of an increase in \To\ from $z > 4$ to $z \sim 3$ is robust, since the colder simulation runs are likely to underestimate the true level of Jeans smoothing, while in the the hotter runs it is likely to be overestimated.  Our analysis should therefore attribute too little of the change in curvature to changes in the instantaneous temperature, which would lead to an underestimate of the change in \To\ with redshift. 

\subsection{Comparison with literature values}

We compare our results with selected \To\ values from the literature in Figure~\ref{fig:T0_with_lit}.  In the left-hand panel we overplot our values with the results of the wavelet analysis performed by \citet{lidz2009}.  The overall agreement is fair, with the exception of $z = 3.4$, where their 2$\sigma$ lower bound significantly exceeds our inferred value of \To\ for $\gamma = 1.3$.  To match their lower limit of 20\,000\,K, we would require a mildly inverted temperature-density relation ($\gamma \sim 0.9$).  While tentative evidence evidence for an inverted temperature-density relation has been found in the flux probability distribution function \citep{becker2007,bolton2008}, it remains a theoretical challenge to simultaneously achieve this over the entire IGM, even during \heii\ reionization \citep{mcquinn2009a}.  We are marginally consistent with the Lidz et al. lower limit at $z = 4.2$.  They note that their errors in this bin are strongly affected by uncertainties in the mean flux, and it is possible that a lower $\tau_{\rm eff}$, such as the one we have used, would bring their \To\ measurement into better agreement with ours.  In the right-hand panel of Figure~\ref{fig:T0_with_lit} we overplot our values with the results of the Voigt profile analysis by \citet{schaye2000}.  The agreement is again fair, although the Schaye et al. errors are only 1$\sigma$.  A careful measurement of \g\ as a function of redshift will be required before our results can be more directly compared to these works, which have both attempted to simultaneously constrain \To\ and \g.

\section{Implications for He~II reionization}\label{sec:implications}

We now discuss  whether our temperature measurements require
that \heii\ reionization occurs within $2 < z < 5$.  Following reionization, gas at the mean density is expected to cool monotonically.  The gas will cool rapidly at first, then asymptotically approach a thermal state where the temperature is set by the shape of the UVB.  In photoionization equilibrium, the thermal asymptote will be of the form $T_{0} \propto (1+z)^{0.53}$ \citep{huihaiman2003}.  If \heii\ is reionized at $z > 5$, then this is the minimum rate at which we would expect the gas to be cooling over $2 < z < 5$.  We show an example thermal asymptote in Figure~\ref{fig:T0}, which has been normalized to intersect our \To\ measurements at $z \ge 4.4$.  This trend falls well below our measured values of \To\ at $z < 4$, even in the case of $\gamma \sim 1.5$, which sets a lower limit for the temperature.  The observed rise in \To\ from $z = 4.4$ to $z \sim 3$ therefore requires an additional source of heating beyond what can be achieved in photoionization equilibrium.  

The expected temperature increase for a parcel of gas due to \heii\ reionization is approximately \citep{furoh2008b}
\begin{equation}
   \Delta T \approx 0.035 \left( \frac{2}{3k_{\rm B}} \right) \langle E \rangle
   \label{eq:DeltaT}
\end{equation}
where $\langle E \rangle$ is the mean energy of the ionizing photons.  In the optically thin limit, the mean energy will be weighted by the ionization cross section, $\sigma_{i} \propto E^{-3}$.  For an ionizing spectrum $J_{\nu} \propto \nu^{-\alpha}$, this will produce $\langle E \rangle_{\rm thin} = 54.4\, (\alpha + 2)^{-1}\, {\rm eV}$, which for a quasar-like spectrum \citep[$\alpha = 1.5$;][]{telfer2002} gives $\Delta T_{\rm thin} \approx 4\,000$\,K.  In the optically thick limit, all ionizing photons will be absorbed and $\langle E \rangle_{\rm thick} = 54.4 \, (\alpha - 1)^{-1}\,{\rm eV}$, or $\Delta T_{\rm thick} \approx 30\,000$\,K.  The increase in \To\ in Figure~\ref{fig:T0} can easily be accommodated within this range.  We emphasize, however, that the temperatures we are measuring are averaged over large volumes.  During an extended reionization process, the first regions to ionize will have time to cool before the overall process completes.  The maximum increase in the globally-averaged \To\ will therefore be smaller than the increased experienced by an individual parcel of gas.  

We leave detailed modeling of the rise in \To\ for  future work. Assuming that the heating of the low-density IGM at these redshifts  is indeed dominated by \heii\ photoheating, however,  we can already formulate  two basic conclusions: (i) that \heii\ reionization occurs at $z < 5$, and (ii) that the process is extended, beginning at $z \ge 4.4$ and potentially extending to $z \sim 3$, consistent with the end of \heii\ reionization inferred from the evolution in the \heii\ mean opacity \citep[][and references therein]{dixon2009, mcquinn2009c}.  The redshift at which $T_0(z)$ begins to decline, as well as the peak value of \To\ will depend primarily on the evolution of \g\ with redshift.

Finally, we note that our values of \To\  at $z \simeq 4-5$ are
consistent with a broad range of hydrogen reionization scenarios.  At $z > 6$,
\To\ is expected to decrease rapidly following \hi\ reionization due
to the combined effects of adiabatic expansion and Compton cooling
\citep{me1994,huignedin1997}.  Even for \hi\ reionization at $z
\gtrsim 10$, however, the asymptotic value of \To\ at $z \simeq 4-5$
is expected to be $\sim$8\,000\,K \citep{huihaiman2003}, which is
consistent with our measurements.  This temperature assumes a
quasar-like background with $\alpha = 1.5$, while a stellar spectrum
with $\alpha = 3.0$ would lower \To\ by $\sim$20 per cent
\citep{huihaiman2003}.  On the other hand, our estimates of the
temperature at $z > 4$ may be too high by roughly the same amount for
the reasons described in Section~\ref{sec:thermal_history}.  Given the
uncertainties in both the measurements and the thermal models,
therefore, we conclude that our temperature measurements at $z > 4$ do
not obviously require a recent ($z < 10$) temperature boost from \hi\
reionization.  This in in contrast to earlier works
\citep{theuns2002b,huihaiman2003}  based on temperatures estimates that were considerably higher than the ones measured here.  We note that the rapid cooling expected after reionization completes means that our measurements are also consistent with \hi\ reionization at $z  < 10$ for a wide variety of reionization redshifts and initial temperature boosts \citep{huihaiman2003,furoh2008b}.  \citet{bolton2010} recently demonstrated that local constraints on the timing of hydrogen reionization can also be placed from temperature measurements in the near-zones of $z \sim 6$ QSOs.  Combining near-zone measurements with our results from the forest should in the future provide a powerful means of studying the history hydrogen reionization. 

\section{Summary}\label{sec:summary}

We have produced new measurements of the IGM temperature over $2.0 \le z \le 4.8$.  These are the most sensitive such measurements to date, and the first in the general IGM at $z >  4.5$.  Our analysis uses a new statistic, the curvature, to characterize the small-scale structure of the \lya\ forest in a manner that is highly sensitive to the IGM temperature while allowing systematic errors to be effectively minimized.  We further focus on measuring the temperature at an optimal overdensity for each redshift, where the temperature is nearly a one-to-one function of the mean absolute curvature.  The temperature at the mean density is then inferred from our \Tdelta\ measurements for different values of the slope for the temperature-density relation.   Our main conclusions are as follows:

\begin{itemize}

\item{The temperature at the optimal overdensity increases from $T(\bar{\Delta}) \approx 9\,000$\,K at $z = 4.4-4.8$, where $\bar{\Delta} = 1.2-1.4$, to $T(\bar{\Delta}) \approx 30\,000$\,K at $z = 2.0$, where $\bar{\Delta} = 5.7$.  This trend is most likely driven in large part by the fact that $\bar{\Delta}$ increases with decreasing redshift, since the temperature of the IGM is generally expected to increase towards higher densities due to adiabatic heating in the collapsing filaments and cooling in the expanding voids.  The \Tdelta\ measurements are the primary result of this study, and can be directly compared with theoretical models of \heii\ reionization.}

\smallskip

\item{We use the maximum slope for the temperature-density relation expected in the absence of \heii\ reionization ($\gamma \simeq 1.4-1.6$, depending on redshift) to set a minimum value of \To\ as a function of redshift.  We also consider the case of $\gamma = 1.3$, motivated by recent \heii\ reionization simulations by \citet{mcquinn2009a}.  In either case (and for any smaller values of \g), we detect an increase in \To\ from $z = 4.4$ to $z = 2.8$.  The minimum increase is $\Delta T_{0} \approx 4\,000$\,K, although the increase may be larger if \g\ decreases over this interval, or if the differing amounts of Jeans smoothing in our simulations have caused us to underestimate the the change in temperature.}  

\smallskip

\item{The temperature evolution from $z \simeq 4.4$ to 2.8 is not consistent with the asymptotic temperature evolution of a photo-ionized IGM expected long after reionization is completed, where \To\ decreases monotonically with redshift \citep{huignedin1997}.   The observed increase in \To\ over this interval requires a substantial injection of additional energy, which we attribute to photoionization heating during an extended period of \heii\ reionization.}

\smallskip

\item{A peak in $T_{0}(z)$ at $z \simeq 2.8$ is inferred for $\gamma \sim 1.5$.  This would be consistent with an end to \heii\ reionization at $z \sim 2.7$, as inferred from \heii\ \lya\ opacity measurements.  However, the location of the peak can change if \g\ evolves with redshift.  Our measurements are consistent with a falloff in \To\  at $z < 2.8$, particularly since \g\ is expected to return to its asymptotic value ($\gamma \sim 1.5$) following the end of \heii\ reionization.}

\smallskip

\item{The value of \To\ we infer from our \Tdelta\ measurements over $z = 4.4-4.8$ is largely independent of \g, since we are already measuring the temperature close to the mean density.  At these redshifts, $T_{0} \approx 8\,000$\,K, which is consistent with \hi\ reionization having completed well before  $z \sim  6$.}
    
\end{itemize}

Two key challenges remain to fully characterizing the thermal evolution of the IGM over this redshift range, at least in a volume-averaged sensed.  The most pressing issue is to establish the shape of the temperature-density relation and its evolution with redshift.  This will clarify the evolution of $T_{0}(z)$ at $z < 4$, and allow us to identify the end of \heii\ reionization based on the redshift at which \To\ peaks.  The second challenge is to determine the amount of Jeans smoothing in the IGM, which can possibly be constrained using lines of sight towards pairs of QSOs \citep{peeples2010b}.  We leave both of these tasks for the future, but note that  the measurements presented here provide a basis for making further progress.  For example, our \Tdelta\ values may be combined with Voigt profile measurements to determine \g\ over $z \sim 2-4$.

Ultimately one would like to characterize not only the evolution of the volume-averaged thermal state of the IGM but also the evolution in its scatter.  A key characteristic of \heii\ reionization should be that growing  \heiii\ bubbles around QSOs introduce spatial fluctuations in both ionization and temperature \citep{sokasian2002, gleser2005, lai2006, paschos2007, furoh2008b}, even if these are partially damped by radiative transfer effects \citep{mcquinn2009a}.  Establishing the volume-averaged evolution of IGM temperatures provides a baseline against which to measure these fluctuations.  The timing and amplitude of the increase in \To, the changes in \g\ and the evolution of the scatter will each be essential ingredients for understanding the role of \heii\ photoheating in the thermal balance of the IGM, and for formulating a more robust picture of when and how \heii\ reionization occurred.

\section*{Acknowledgements}

The authors would like to thank Bryan Penprase for reducing much of the low-redshift HIRES data, and the anonymous referee for their helpful suggestions.  We also wish to recognize and acknowledge the very significant cultural role and reverence that the summit of Mauna Kea has always had within the indigenous Hawaiian community.  We are most fortunate to have the opportunity to conduct observations from this mountain.  The hydrodynamical simulations used in this work were performed using the Darwin Supercomputer of the University of Cambridge High Performance Computing Service (http://www.hpc.cam.ac.uk/), provided by Dell Inc. using Strategic Research Infrastructure Funding from the Higher Education Funding Council for England.  GB acknowledges financial support from the Kavli foundation.  JB has been supported by an ARC Australian postdoctoral fellowship (DP0984947).  WS has been supported by the National Science Foundation through grant AST-0606868.

\bibliographystyle{apj}
\bibliography{curvature_refs}

\appendix

\section{Numerical Convergence Tests}

\begin{figure*}
   \centering
   \begin{minipage}{\textwidth}
   \begin{center}
   \includegraphics[width=0.75\textwidth]{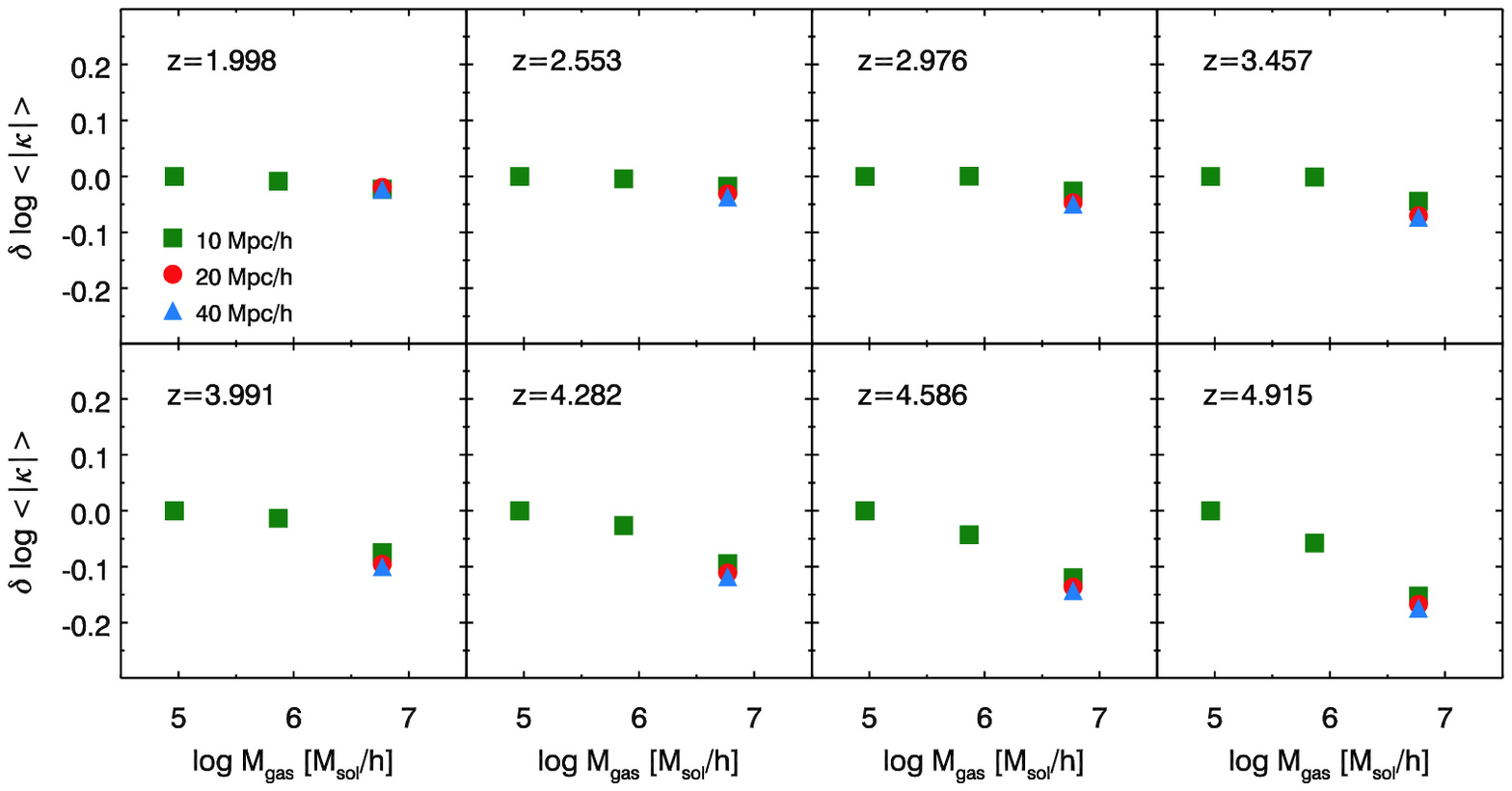}
   \vspace{-0.05in}
   \caption{Convergence of the mean absolute curvature with box size and resolution.  Curvature values were computed from spectra without noise, using the same method as was applied to the real data.  The change in \mac\ is measured with respect to the highest-resolution case.  We are only marginally converged with mass resolution at the high redshift end, but the likely change in \mac\ gained from further increasing the number of particles would only correspond to a small change in the measured temperature (see text).  We do apply a box size correction to the curvature, although it is small.}
   \label{fig:curv_convergence}
   \end{center}
   \end{minipage}
\end{figure*}

Capturing the small-scale structure in the high-redshift \lya\ forest poses a significant challenge for cosmological simulations.  As discussed by \citet{boltonbecker2009}, most of the transmission in the forest at $z \gtrsim 5$ comes from voids, which are less well resolved in SPH simulations than the filaments that dominate the characteristics of the forest at $z \sim 2-3$.    Ideally, one would use both a large simulation box ($l \gtrsim 40$\,\hinv\,Mpc), and high resolution ($M_{\rm gas} \lesssim 10^5\,{h^{-1}}\,{\rm M}_\odot$).  Computational constraints, however, required us to focus on obtaining the necessary mass resolution, which was found to be a more stringent requirement for the curvature.  To check our convergence with mass resolution and box size, we ran a set of comparison simulations (runs R1 -- R4).  In these runs we used the same thermal history as in run C15, but varied the box size and mass resolution.  Compared to run C15, which uses $L = 10$\,\hinv\,Mpc, $n = 2 \times 512^3$~particles and $M_{\rm gas} = 9.2 \times 10^{4}\,h^{-1}\,M_{\odot}$, the mass resolution was lowered by up to a factor of 64, and the box size was increased by up to a factor of four in length.  The optical depths in each simulation were rescaled such that the mean fluxes were equal after the spectra were re-normalized in 10\,\hinv\,Mpc sections (Section~\ref{sec:mean_flux}).  The curvature was then computed using the procedure outlined in Section~\ref{sec:method_summary}, with the modification that noise-free spectra were used and no b-spline fits were performed.  

The convergence results are shown in Figure~\ref{fig:curv_convergence}, where we plot the change in \logmac\ with respect to run C15.  We are nearly converged with box size at all redshifts.    At a fixed resolution, \logmac\ always decreases by less than 0.03 when increasing the box size from 10 to 40\,h\,Mpc.  For the curvature values measured in the data, this corresponds to a temperature difference of less than 900\,K.  A box size correction calculated from these results was applied to the simulation curvature values when fitting the temperatures.  At $z < 4$ we are well converged with mass resolution, although the convergence is less clear at higher redshifts.  At $z = 4.915$, the increase in \logmac\ when increasing the number of particles from $2 \times 256^3$ to $2 \times 512^3$ is 0.058.  We can estimate the gains from an additional factor of eight in mass resolution by comparing to $z = 3.991$, which has a similar change in \logmac\ when increasing the number of particles from $2 \times 128^3$ to $2 \times 256^3$.  In that case, further increasing the mass resolution by a factor of eight increased \logmac\ by only 0.014, which at $z = 4.915$ would correspond to a change in temperature of roughly 500\,K.  While we do not apply a mass resolution correction to our results, therefore, we can be reasonably confident that the gains from going to higher resolution would be small, and that they would only be significant our highest redshift bins where the correction is likely to be smaller than the errors.

\end{document}